\documentclass[12pt]{article}

\usepackage{amsmath}
\usepackage{amssymb}
\usepackage{graphicx} 
\usepackage{hyperref}
\usepackage{listings}
\usepackage{multirow}
\usepackage{multicol}
\usepackage{wrapfig}
\usepackage[dvipsnames]{xcolor}

\usepackage{amscd}
\newtheorem{dfn}{Definition}[section]
\newtheorem{thm}{Theorem}[section]

\usepackage{fdsymbol}
\newcommand{\mysquare}[1]{\textcolor{#1}{$\medblacksquare$}}
\newcommand{\mydiamond}[1]{\textcolor{#1}{$\medblackdiamond$}}
\newcommand{\mytriangle}[1]{\textcolor{#1}{$\medblacktriangleup$}}
\newcommand{\mycircle}[1]{\textcolor{#1}{$\medblackcircle$}}

\newcommand{\tcirc}[1]{\raisebox{.5pt}{\textcircled{\raisebox{-.9pt} {#1}}}}

\begin{document}
\title{On the evolution of Betti curves in the Cosmic web}
\author{Vitalii Tymchyshyn$^{1,2}$,  Maksym Tsizh$^{3,4}$, Franco Vazza$^{3,5,6}$, Marco Baldi$^{3,7,8}$}
\maketitle
\centerline{\it $^1$Bogolyubov Institute for Theoretical Physics}
\centerline{\it Metrolohichna 14-b, UA-02000 Kyiv, Ukraine}
\medskip
\centerline{\it $^2$Data Science Center, Kyiv Academic University, UA-03142 Kyiv, Ukraine}
\medskip
\centerline{\it $^3$Dipartimento di Fisica e Astronomia, Universitá di Bologna}
\centerline{\it Via Gobetti 92/3, 40121, Bologna, Italy}
\medskip
\centerline{\it $^4$ Astronomical Observatory of Ivan Franko National University of Lviv,}
\centerline{\it Kyryla i Methodia str. 8, Lviv, 79005, Ukraine }
\medskip
\centerline{\it$^5$ INAF-Istituto di Radio Astronomia }
\centerline{\it Via Gobetti 101, 40129 Bologna, Italy}
\medskip
\centerline{\it$^6$ Hamburger Sternwarte, Universitat Hamburg }
\centerline{\it Gojenbergsweg 112, 41029 Hamburg, Germany}
\medskip
\centerline{\it$^7$ INAF - Osservatorio Astronomico di Bologna}
\centerline{\it Via Piero Gobetti 93/3, 40129 Bologna BO, Italy}
\medskip
\centerline{\it$^8$ INFN - Istituto Nazionale di Fisica Nucleare}
\centerline{\it Sezione di Bologna, Viale Berti Pichat 6/2, 40127 Bologna BO, Italy}

\begin{abstract}
In this work, we study the evolution of Betti curves obtained by persistent-homological analysis of point clouds formed by halos in different cosmological $N$-body simulations. We show that they can be approximated with a scaled log-normal distribution function with reasonable precision. Our analysis shows that the shapes and maximums of Betti curves exhibit dependence on the mass range of the selected subpopulation of halos. Still, at the same time, the resolution of a simulation does not play any significant role, provided that the mass distribution of simulated halos is complete down to a given mass scale. Besides, we study how Betti curves change with the evolution of the Universe, i.e., their dependence on redshift. Sampling subpopulations of halos within certain mass ranges up to redshift $z=2.5$ yields a surprisingly small difference between corresponding Betti curves. We propose that this may be an indicator of the existence of a new specific topological invariant in the structure of the Universe.
\end{abstract}
\section{Introduction}
\label{sec:Intro}

The topological data analysis of the Cosmic web is a rapidly developing field of study. The foundation in this field was set with early works by A. Hamilton \cite{Hamilton86}, V. Sahni \cite{Sahni98}, Sheth \cite{Sheth03} and their colleagues, in which they used notions of curvature of the density field, Minkowski functionals, and genus to characterize the large scale structure distribution of matter. Another early result is the algorithm of large scale structure identification and matter density recovery from discrete (point cloud) data based on the Delaunay tessellation field estimator (DTFE) and persistent topology structures, DisPerSE, developed by T. Sousbie and colleagues \cite{Sousbie1}, \cite{Sousbie2}. More complex methods were developed in the first decades of this century.

The application of persistent homology is one of the most important and interesting latest advancements in this direction. Information on the persistent homology of the given point cloud can be aggregated into Betti curves (sometimes dubbed as Betti functions) that show the dependence of Betti numbers on the persistence parameter. There are two general ways to apply this tool in cosmology and large scale structures studies. The first one uses Betti curves directly for characterizing certain $3$-($2$-)dimensional scalar fields: matter density, ionization maps of HI, etc. In this case, the persistence is measured on the scale of the intensity contrast parameter of the field. Thus, one calculates Betti numbers for manifolds generated by sub-(super-)level sets of the field. 
The second way is the utilization of Betti curves of discrete data, i.e., point clouds, by building certain series of simplicial complexes (e.g., \u{C}ech, Vietoris-Ripps, Witness, etc. complexes, in general, called $\alpha$-complex) with linking length that plays the role of a persistence parameter (also called filtration parameter) in that case. Typically, $\alpha$-complex is used as computationally efficient. The point cloud under consideration in large scale studies is usually a population of dark matter halos or galaxies in simulation or real Universe. We will be interested in the latter type of Betti curves application, but let us first overview both application methods.

The application to continuous fields is often exploited for discovering non-Gaussianity in distribution and for comparison with random fields in general. It was performed for both two-\cite{Feldbrugge19} and three-\cite{Park13} dimensional continuous random fields. Among other usages, we can mention studying the topology of the reionization bubbles in \cite{Giri21, Elbers19, Elbers23}. All of these works address the problem of studying the topology structure of a map of ionized regions during the reionization epoch - and include analysis of their Betti curves. The cosmic microwave background (CMB) radiation temperature map was also subjected to persistent homology analysis \cite{Cole18}. The authors demonstrate that such Betti curves are sensitive to non-Gaussianity in the CMB spectrum. \cite{Heydenreich21} have performed an excellent analysis of persistent Betti numbers of gravitational lensing maps, proposing persistence homology methods to resolve modern $S_8$ tension in cosmology. This was probably the first quantitative application of a persistent homology instrument for restricting cosmological parameters. An alternative approach when working with continuous fields (of density) is to use Morse theory \cite{Codis18} and Morse-Smale complexes\cite{Shivashankar16} which can also yield a skeleton and elements of large scale structure, such as filaments. The last two works are examples of how one can reconstruct the continuous field from a point cloud data. Reconstruction through DTFE from the distribution of particles of cosmological simulation with the following analysis of Betti curves was also performed in \cite{Pranav16}. It has shown that Betti curves can ``feel'' the mean distance between particles, the concentration of points in a point cloud, and the evolution of particle distributions.

Studies of cosmic $\alpha$-complexes yielded many interesting results, too. Pioneers here were "the Groningen group," which was started by Rien van de Weygaert and his colleagues. Not only did they perform a large portion of the applications of the first kind that we just cited, but also, in 2011, they were the first who show that Betti curves are somewhat sensitive to dark energy \cite{Weygaert11}. Then, they performed a more fundamental persistence homology analysis of the Cosmic web in 2013 \cite{Weygaert13}. It also includes the analysis of the evolution of Betti curves of the Cosmic web. In their first attempt, however, they did not consider a phenomenon they would later call a "topological bias": the topological properties of manifold induced by an $\alpha$-complex is tightly connected to the mass range of halos included in the considered web. Later on, \cite{Xu19} showed that Betti curves are sensitive to such non-obvious cosmological parameters as the sum of neutrino masses. The  ``Groningen group'' continued their work on deep TDA analysis of the Cosmic web in 2021: they studied \cite{Wilding21} the persistence of Betti numbers for superlevel sets filtration 3D-density map, with density contrast as persistence parameter. They've shown these curves' evolution and evaluated how their approximations' parameters evolve. Next year, they discovered \cite{Bermejo22} the phenomenon of "topological bias" by analyzing subpopulations of halos of different mass ranges. They show how different the Betti curves are for these subpopulations.

Finally, \cite{Ouellette23} have recently presented the deep analysis and comparison of Betti curves of galaxy point clouds in simulations and observations. With the help of Betti curves, they have concluded that the topology of the galaxy distribution is significantly different from the topology of the dark matter halo distribution. 

In the current contribution, we analyze Betti curves generated by filtration of $\alpha$-complexes built on a point cloud of halos of three different dark matter-only simulations. We show that Betti curves obtained this way can be approximated as scaled log-normal distribution with reasonable precision. Moreover, we track their evolution through cosmic time and elaborate on the factors that might impact them. We will show that sampling the subpopulation of halos within a certain mass range up to redshift $z=2.5$ yields a surprisingly small difference between corresponding Betti curves, which we proposed is an indicator of the existence of a certain topological invariant of the Universe. The comparison with random point clouds supports this idea, and the stability of Betti curves that we derived in this work is thus not an artifact, but it stems from the physics of simulated structure formation. 

We organized this work as follows: in sec.\ref{sec:betticurves} we will briefly explain the notion of Betti numbers and Betti curves prescribed to a given point cloud. The next sec.\ref{sec:simulations} is devoted to the details of the cosmological simulation we used in the current contribution. Then, in sec.\ref{sec:results}, we share the results of computing Betti curves of different simulations and comparing them to each other. The last sec.\ref{sec:conlusions} contains conclusions. 

\section{Betti curves}
\label{sec:betticurves}
When exploring the Cosmic web, we are essentially dealing with a large point cloud.
The challenge in handling this point cloud lies not only in the humongous amount of data but also in the fact that, in addition to physical laws, it encompasses a number of random effects.
This leaves us pondering whether a certain observed pattern results from natural law or if it simply occurred by chance. To address the problem, one can refrain from focusing on individual points and instead extract generalized information about their distribution. This approach will yield a manageable amount of data and, hopefully, reduce the randomness in favor of regularity.

One of the possible approaches is to endow each point with a sphere of radius $r$ centered at the point and consider the union of all these balls as a manifold. This manifold will possess specific topological properties that can be represented numerically and utilized in subsequent analysis. As such, one uses Betti numbers\,--- ranks of appropriate homology groups of the given manifold. For simplicity, one can envision the $0$-th Betti number as the count of connected components, the $1$-st as the count of one-dimensional or ``circular'' holes, the $2$-nd as the count of two-dimensional ``voids'' or ``cavities,'' and so forth.

This notion is quite convenient as we can associate Betti numbers with cosmological structures, even though this connection is somewhat indirect: $0$-th with clusters, $1$-th with cycles and tunnels formed by cosmic filaments, and $2$ with cosmic voids formed by cosmic sheets. Moreover, representation through Betti numbers is quite stable with respect to translation, rotation, and small variation of points' positions, which is a desired property. In our work, we extract topological features using the so-called $\alpha$-complex without imposing any restrictions on the parameter $\alpha$. One may be concerned that it is defined in terms of simplices and Delaunay triangulation but not intersecting balls. Though, without any restriction on $\alpha$, it is equivalent to the \u{C}ech complex (but much smaller, which is beneficial for calculation), and \u{C}ech complex in its turn is homotopically equivalent to the ``intersecting spheres'' point of view by the classic ``Nerve Theorem'' \cite{alexandroff1928}. Thus, both views are equivalent, but ``intersecting spheres'' are easier to grasp. 

However, the choice of the radius $r$ to use is debatable. A potential solution is to consider all possible values of $r$ and monitor the number of topological features, i.e., Betti numbers, as the manifold changes with varying $r$. Plotting Betti number vs $r$ we get a so-called ``Betti curve.'' 
By considering all possible values of $r$ with some small step, one gets rid of bringing a redundant parameter to the quantitative method of studying the Cosmic web. The benefit of such a parameterless approach, as we will see below, is the possibility of finding natural, inherent scales for different structure types.

It is worth noting that one can work with individual topological features as well, e.g., in \cite{Tsizh23} we apply persistence homology methods to a small-scale Cosmic web. In this work, we restrict ourselves to only 1- and 2-dimensional homology features of the $\alpha$-complexes manifolds. One can interpret them as filament loops and voids correspondingly, remembering, however, that cosmic filaments and cosmic voids are not to be confused with those.

We provide strict algebraic definitions of a simplicial complex, its filtration, homology groups, Betti number and curves, and theorems related to their properties in Appendix~\ref{sec:appendixa}. 
We refer the reader to more thorough, copious, and complete works of H. Edelsbrunner\cite{Edelsbrunner02}, A. Zomorodian \cite{zomorodian09}, \cite{zomorodian12}, and G. Carlsson \cite{carlsson04} \cite{carlsson09} where the rigorous notion of complex filtration was established, as well as algorithms provided for its computation and application. See also work by L. Wasserman \cite{Wasserman18} for a more compact modern review of topological data analysis instruments.

\section{Cosmological simulations}
\label{sec:simulations}

We have used three large and widely used cosmological simulations to study the Betti curves generated by halos: Bolshoi\cite{bolshoi}, Multidark\cite{multidark}, and the LCDM subset of the DUSTGRAIN-pathfinder simulations \cite{Giocoli:2018gqh} that was developed for the Higher-Order Weak Lensing Statistics (HOWLS) project \cite{Euclid:2023uha}. All of them are dark matter-only simulations and were produced with GADGET2/GADGET3 code. These simulation share almost identical cosmological model, however, they significantly differ in their mass and spacial resolution, with Bolshoi being the most resolved, and HOWLS being the least resolved one. In all three simulations, halos were identified by the Friend-of-Friend (FoF) algorithm. The values of the linking length in terms of mean particle separation, gravitational softening parameter, and other simulation parameters, as well as cosmological parameters for all three simulations are given in table~\ref{tab:simulations}. We considered halos only for redshifts $z \leq 2.5$, as only starting from such redshifts all of these simulated universes contain a sufficiently large number of halos for our analysis. When using Bolshoi simulation data, we further restricted ourselves to only 2 heaviest millions of the 10 million (at $z=0$) halos since processing such large halo numbers is highly time-consuming.

\begin{figure*}[!hbt]
\begin{center}
\begin{minipage}{0.8\textwidth}
    \includegraphics[width=\textwidth]{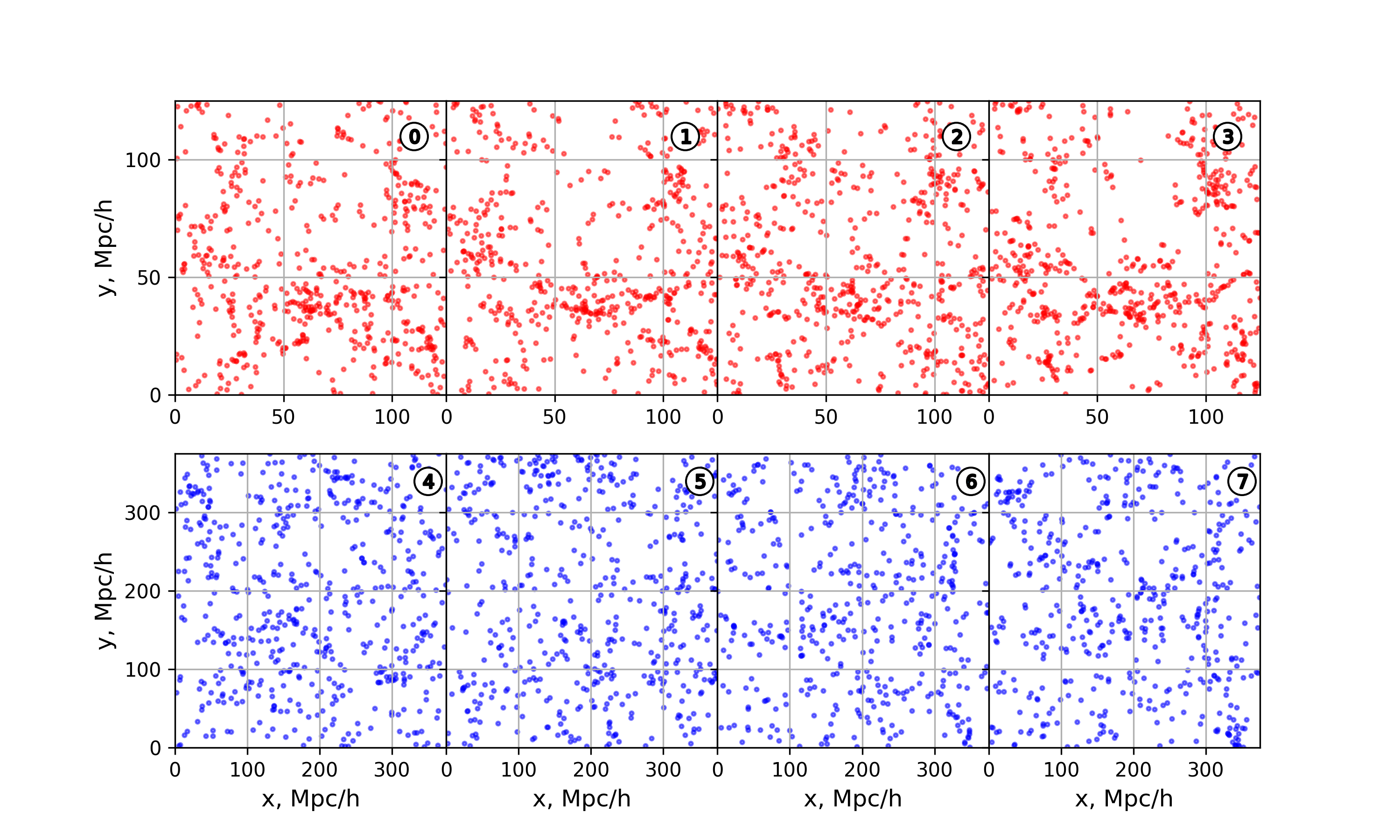}
\end{minipage}\hspace{-1cm}
\begin{minipage}{0.2\textwidth}
\footnotesize
\begin{tabular}{rll}
\# & $\ln(M/M_\odot)$ & $z$\\
\hline
\tcirc{0} & $11.0\dots 11.5$ & $0$\\
\tcirc{1} & $11.5\dots 12.0$ & $0$\\
\tcirc{2} & $11.0\dots 11.5$ & $2$\\
\tcirc{3} & $11.5\dots 12.0$ & $2$\\
\tcirc{4} & $12.0\dots 12.5$ & $0$\\
\tcirc{5} & $12.5\dots 13.0$ & $0$\\
\tcirc{6} & $12.0\dots 12.5$ & $2$\\
\tcirc{7} & $12.5\dots 13.0$ & $2$\\
\end{tabular}
\end{minipage}
\caption{Spatial distribution of halos. Upper row\,--- Bolshoi, lower\,--- HOWLS. The depth of slice is $10$~Mpc/h for Bolshoi and $30$~Mpc/h for HOWLS}
\label{fig:slices}
\end{center}
\end{figure*}

\begin{figure*}[!htb]
\begin{center}
\includegraphics[width=0.48\textwidth]{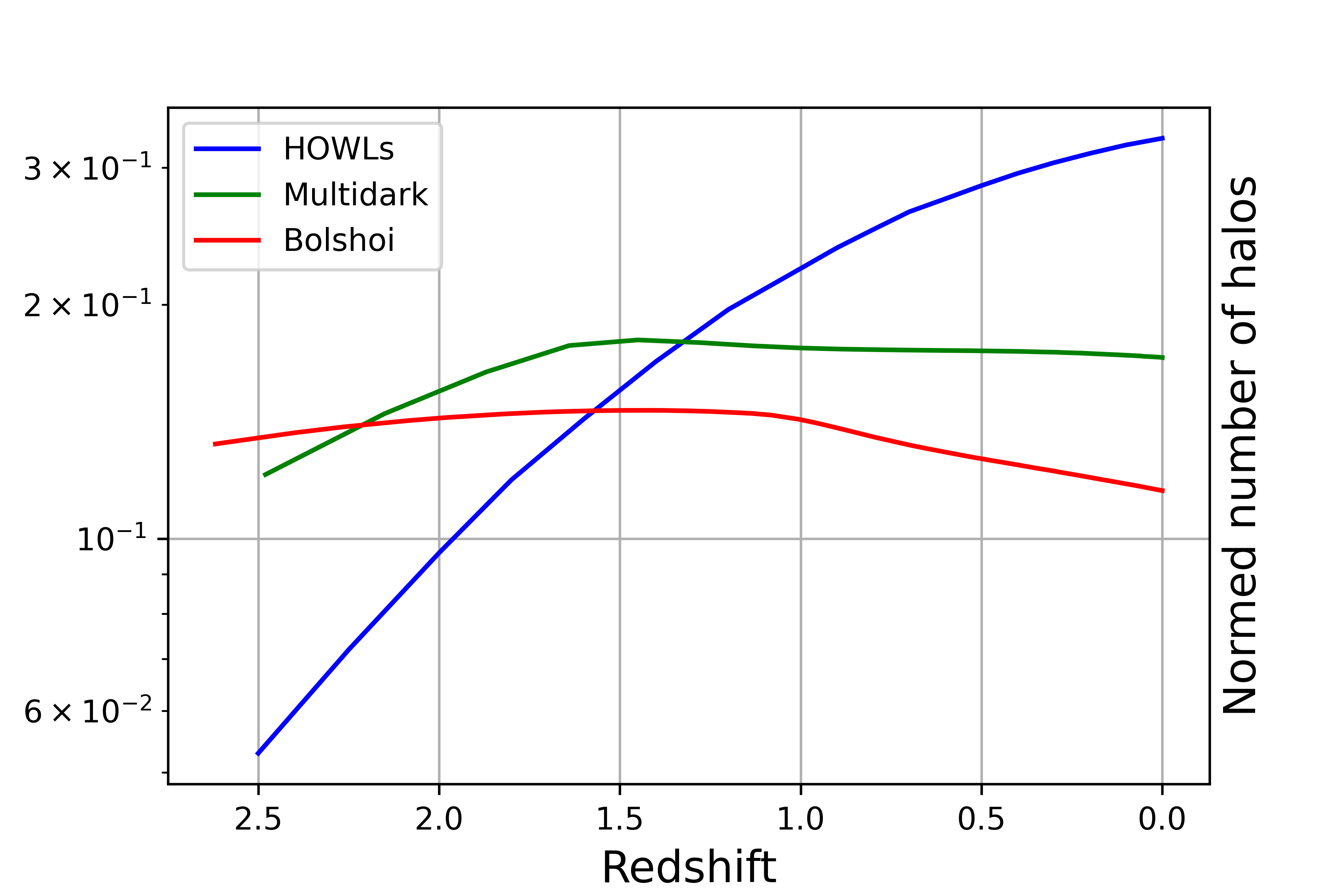}
\includegraphics[width=0.48\textwidth]
{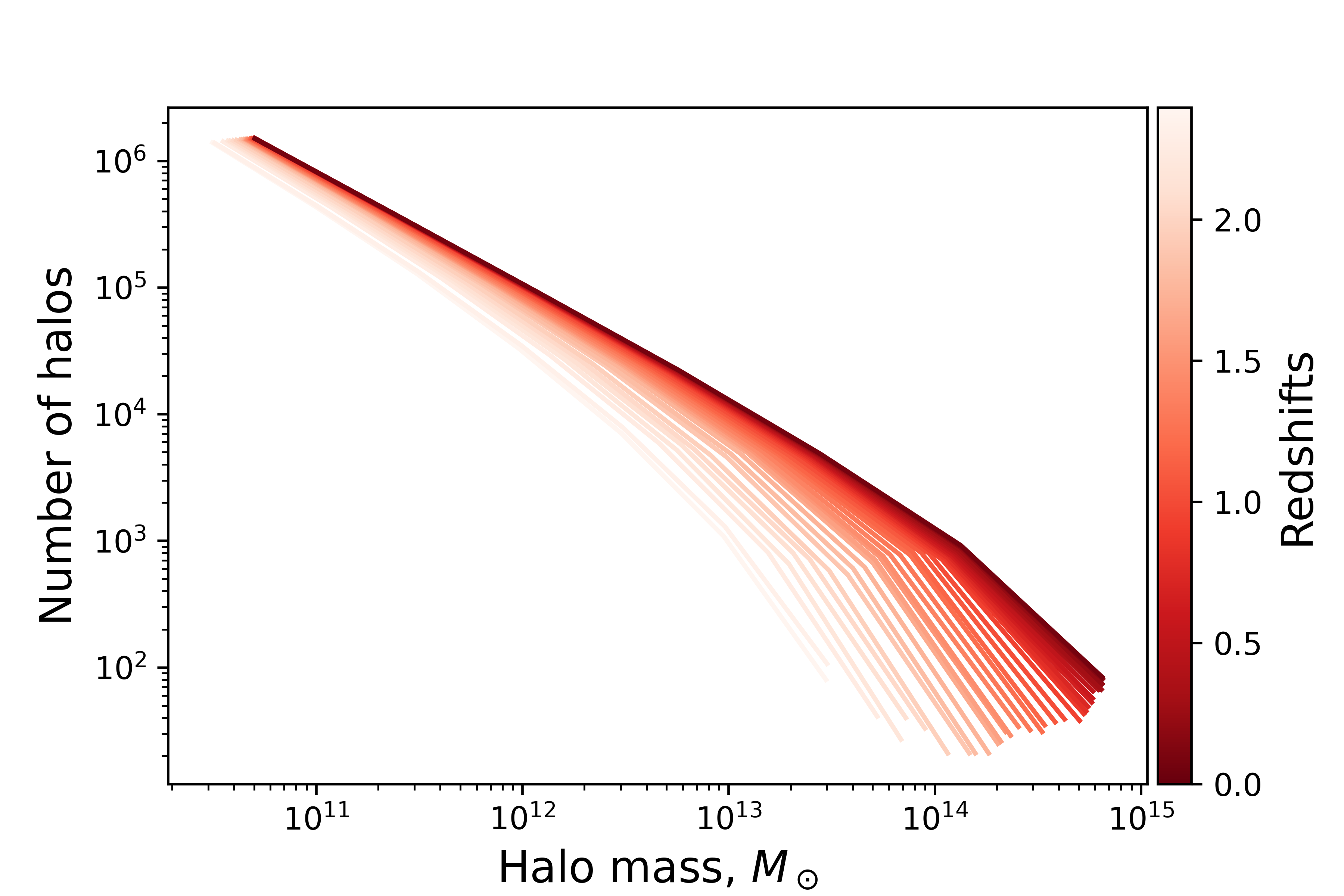}
\includegraphics[width=0.48\textwidth]
{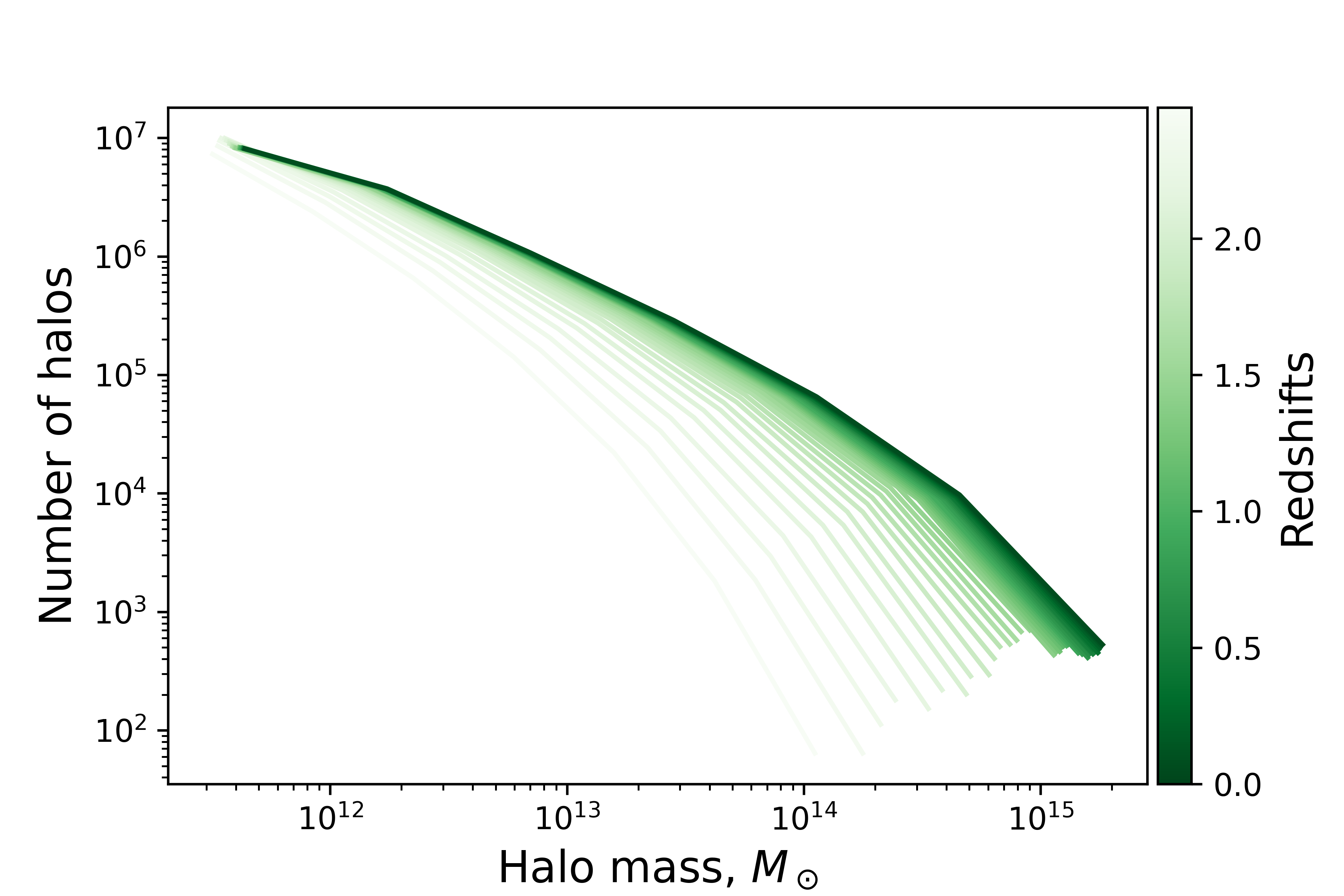}
\includegraphics[width=0.48\textwidth]
{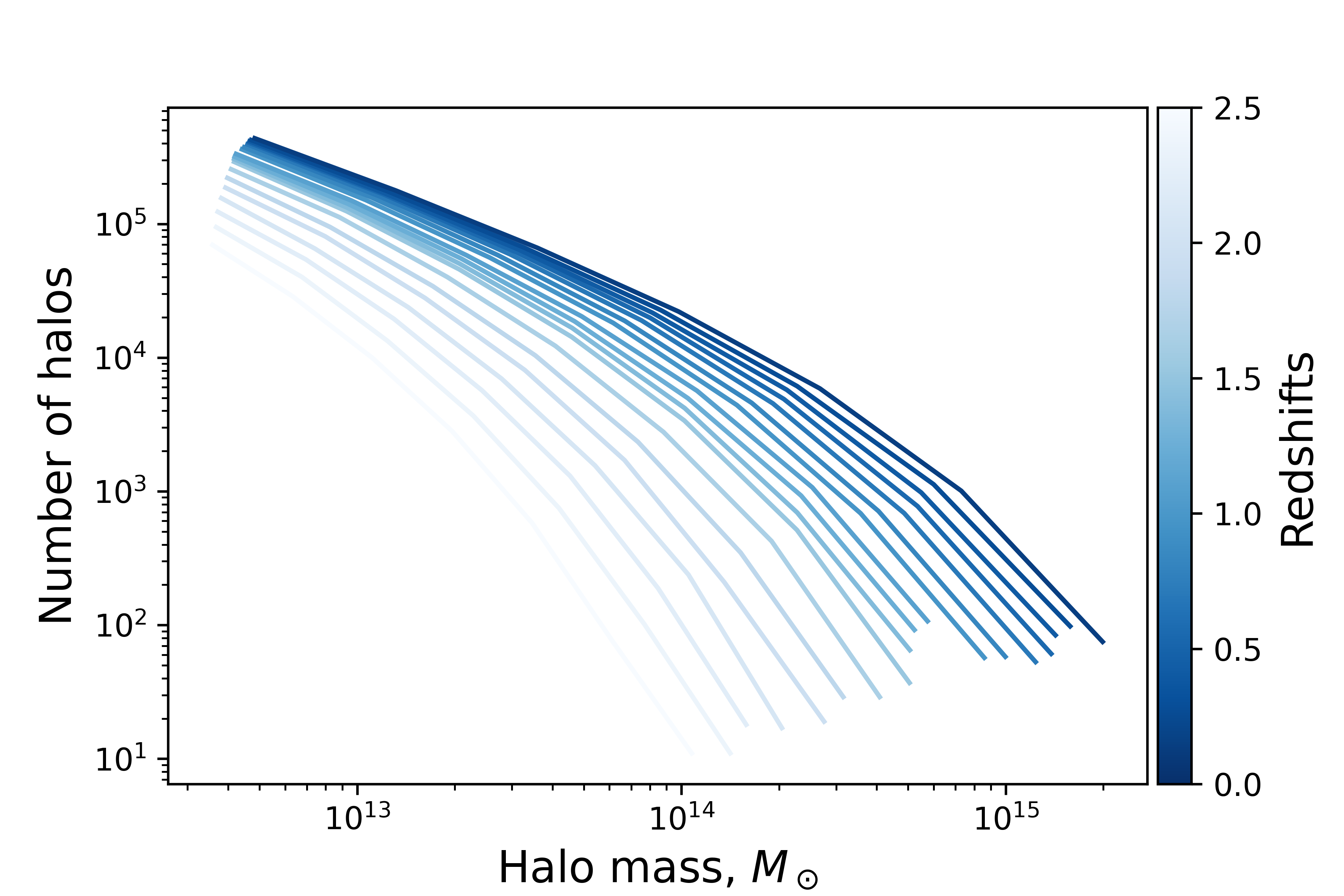}
\caption{Top-left panel: The normalized total number of halos change in studied simulation with cosmological evolution. Other panels: Evolution of halo mass function of total halo population in Bolshoi (top-right), Multidark (bottom-left), and HOWLS (bottom-right) simulations}
\label{hmf}
\end{center}
\end{figure*}

The key difference between simulations in mass (and spatial) resolution leads to more subtle divergences, which we would like to point out. In fig.\ref{hmf}, one can find how the normalized total number of halos changes through cosmological time for each simulation. It is noticeable that the number of halos in HOWLS increases monotonically through the whole cosmic period considered here, while for Multidark and Bolshoi the peaks in the number of halos occur at earlier redshifts. This can be explained by the fact that the low-mass halos dominate the simulations with the highest resolution. Indeed, it was well-established more than a decade ago that small halos form much earlier than larger ones, and their number starts to decline when they merge into more massive halos (see Fig. 2 in \cite{Lukic07}). Nevertheless, they prevail at low redshifts. Hence, the behavior of the total number of halos for simulation with higher resolution is determined by the number of the lightest halos. As we will see below, the mass range of the considered subpopulation strongly impacts the shape and maximum of the correspondent Betti curve. This is why we underline such important features of the halo populations of different simulations. The halo mass functions of all simulations are given in fig \ref{hmf}. Note how similar the halos' mass functions of different simulations are. They do not display the distinguishing features of mass distribution discussed above.

We depict slices of halo spatial distribution for different redshifts and mass ranges for considered simulations in figure~ \ref{fig:slices}. Halo correlation functions are given in figure~ \ref{fig:corrf}. They can help the reader to access the so-called "clustering" effect on large scale structure: the tendency of heavier halos to cluster more in comparison to lighter ones. This becomes important regarding the "topological bias" effect on Betti curves, which will be discussed below.
\begin{table}[]
\begin{tabular}{lllllllll}
\hline
Simulation & Volume, & Cosmology & $\Omega_{M}$& $\Omega_\Lambda$ &$\Omega_b$&$\sigma_8$ &$n_s$  &$H_0$,\\
Name/Units & $(\text{Mpc/h})^3$ &  & &  && & &$\rm km/s/Mpc$\\
\hline
Bolshoi    &  $250^3$            &   WMAP5\cite{wmap5}         & $0.27$& $0.73$ &$0.0469$ & $0.82$& $0.95$&$70$\\
Multidark  &  $1000^3$            &  WMAP5           & $0.27$& $0.73$ &$0.0469$ & $0.82$& $0.95$&$70$\\
HOWLS      &  $750^3$            &  Planck15\cite{planck15}          & $0.31345$& $0.68655$ &$0.0481$& $0.847$& $0.9658$&$67.31$
\end{tabular}
\begin{tabular}{llllllll}
\hline
Simulation & $N_{particles}$ & $m_{particle},$  & Grav.  & FoF  &Total $N_{halos}$&Minimal \# of &$z_{init}$ \\
Name/Units & --- & $h^{1} M_\odot$ & softening  & resol.  & at z=0 & particles in halo & ---\\ 
\hline
Bolshoi    &  $2048^3$          &  0.135               &   1  kpc/h                     &   0.17              & $1.01\cdot 10^7$ & 100 & 80\\
Multidark  & $2048^3$          &  0.87               & 7  kpc/h                       & 0.17               & $1.3\cdot 10^7$  & 100 & 65\\
HOWLS      &  $768^3$           &  5.8               &  24 kpc/h                        &   0.16         
 & $7.3\cdot 10^5$ & 32 & 99 \end{tabular}
\caption{Cosmological parameters of the simulations used in this work.}
\label{tab:simulations}
\end{table}

We have computed Betti curves for each of the universes at given redshift $z$, for all halos, and for several specific mass ranges. We then compare the latter pairwise (Bolshoi vs. Multidark and Multidark vs. HOWLS) to avoid comparing the ones with an enormous difference in mass/spatial resolution.

\section{Results}
\label{sec:results}
All of the results were obtained with routines written\footnote{One can find our code in this repository: \href{https://github.com/mtsizh/betti_curves_evolution}{github.com/mtsizh/betti$\_$curves$\_$evolution}} in \texttt{Python}, using a specialized library for persistent homology computations --- \texttt{Gudhi}. It is well known that the properties of Betti curves directly depend on the concentration of points in a point cloud. Therefore, we shall compare Betti curves computed on halo point clouds with the same concentration of halos. In our case, we will do it for the same (sub)volumes of simulations in which we sample an equal number of halos.  

\subsection{Raw Betti curves}
\begin{figure}[!htb]
\begin{center}
\includegraphics[width=0.48\textwidth]
{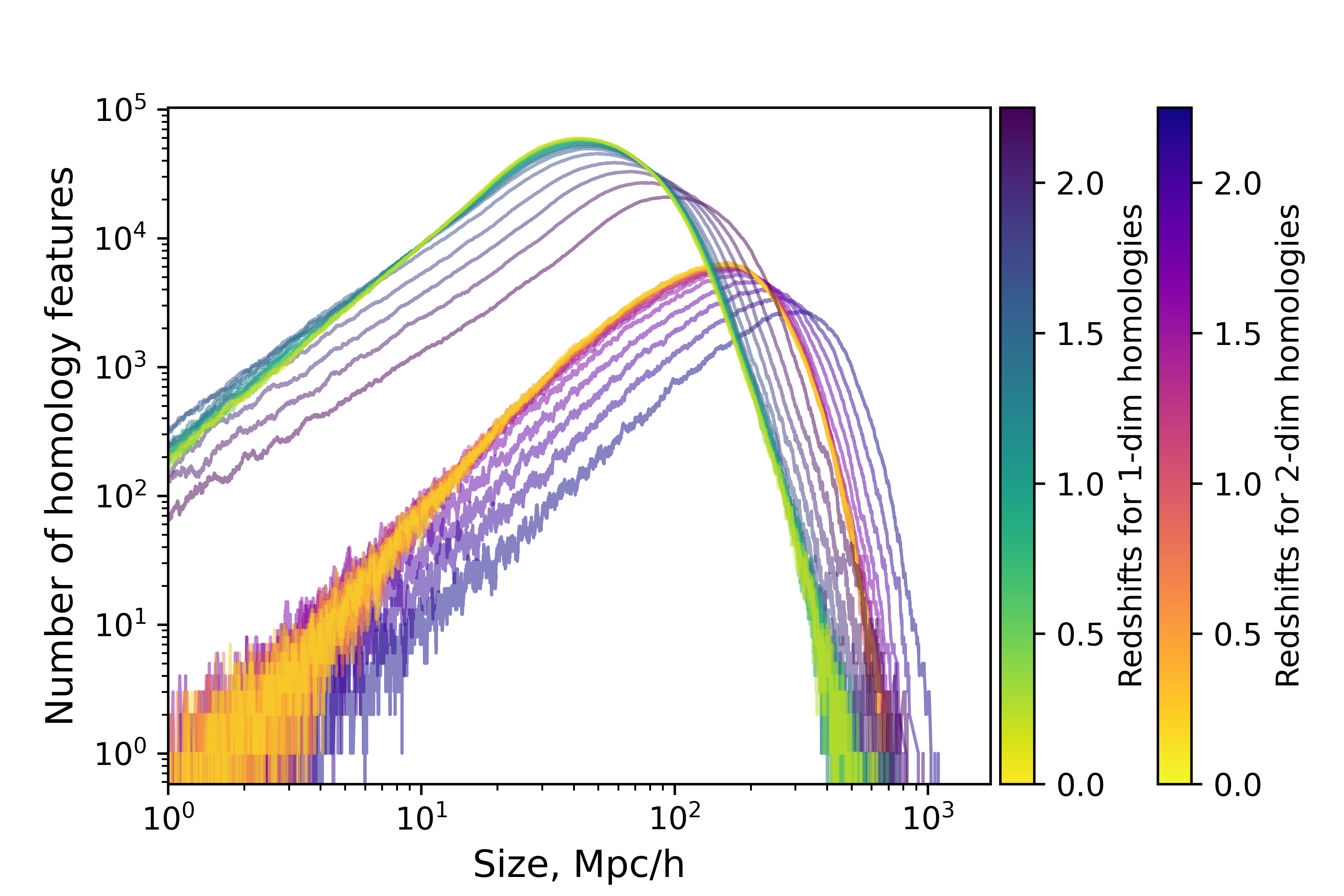}
\includegraphics[width=0.48\textwidth]
{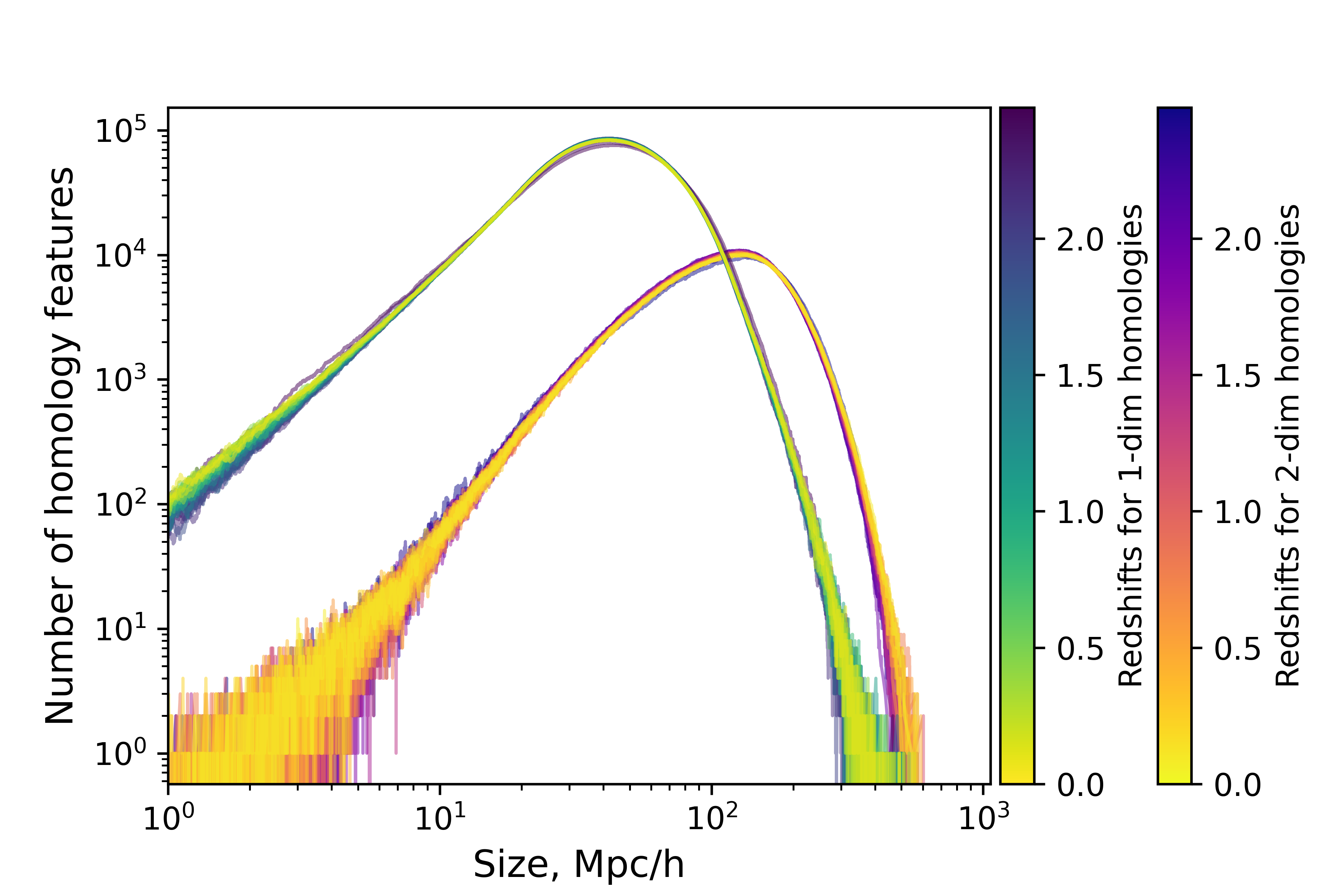}
\caption{Evolution of Betti curves for HOWLS (left) and Multidark (right). Both $1$- and $2$- dimensional features are in one picture: peaks of $1$- dimensional features Betti curves are on lower radii (sizes) than of $1$- dimensional ones.}
\label{betticurvesall}
\end{center}
\end{figure}

First, let us compare the evolution of Betti curves on cosmological time for complete halo populations. Fig.\ref{betticurvesall} depicts the evolution for HOWLS and Multidark simulations in terms of $1$- and $2$-homologies. Recall that for each value of filtration parameter $\alpha$ (analogous to ball radius $r$ for \u{C}ech complex) Betti curve shows how many topological features (``holes'' of certain dimensions) are present in the corresponding manifold. First, it can be noticed that the evolution of the curves is more evident for the HOWLS simulation, which also has a larger dispersion in the peak values. Next, we notice that each dimension has its own span of persistence scales, being larger for $2$-dimensional features.

The ``noise'' in these graphs has a simple explanation: the Betti curve changes its value each time a feature is born or dies when increasing the filtration parameter. Since many features have a very short ``lifespan,'' one can observe abrupt changes in value that can be interpreted as noise, while the general tendency is determined with more persistent features. The noise seems to be larger in amplitude near the origin of the graph, as we have used a log-log scale. 

\begin{figure*}[!htb]
\begin{center}
\begin{tabular}{cc}
\includegraphics[width=0.48\textwidth]
{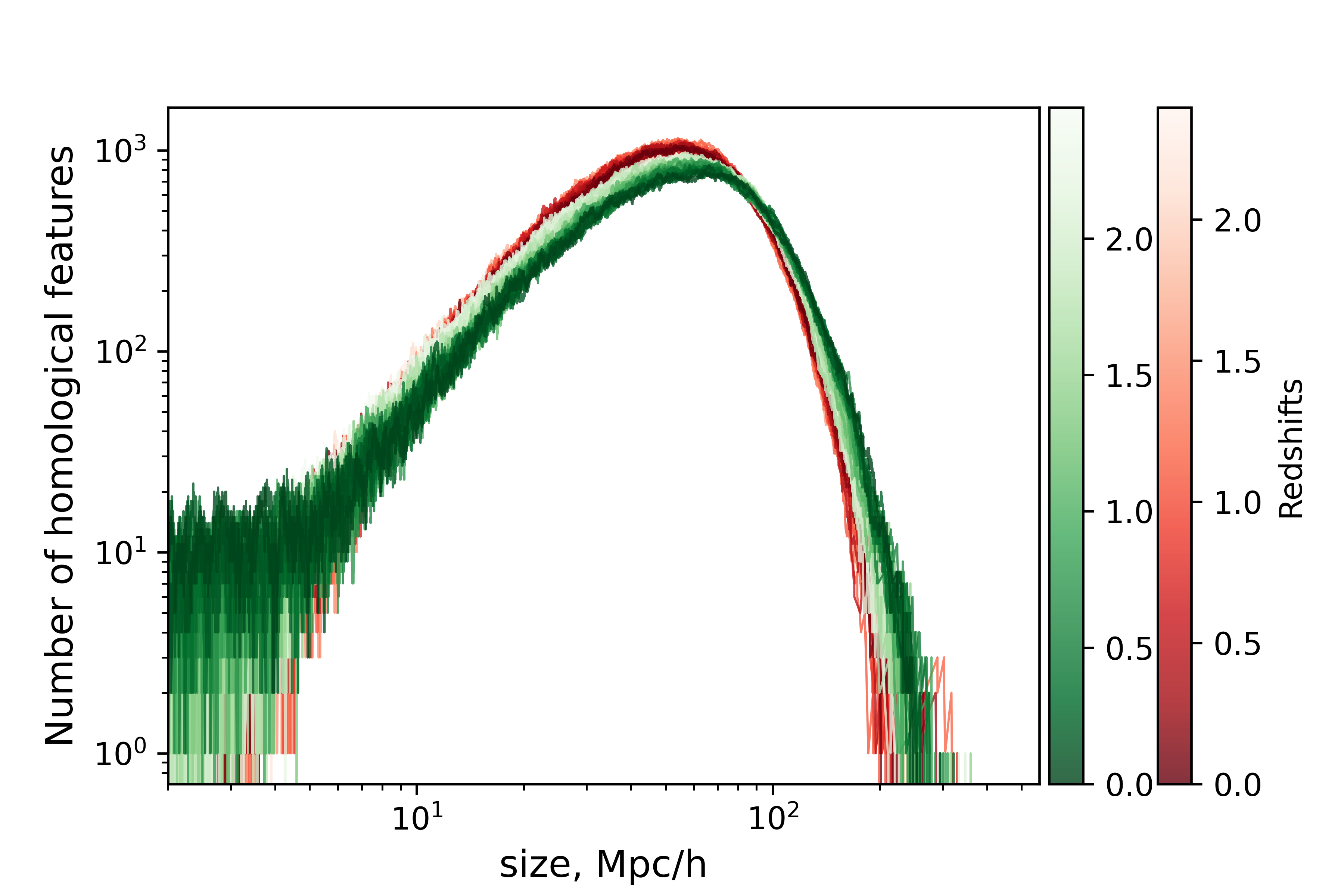} &
\includegraphics[width=0.48\textwidth]
{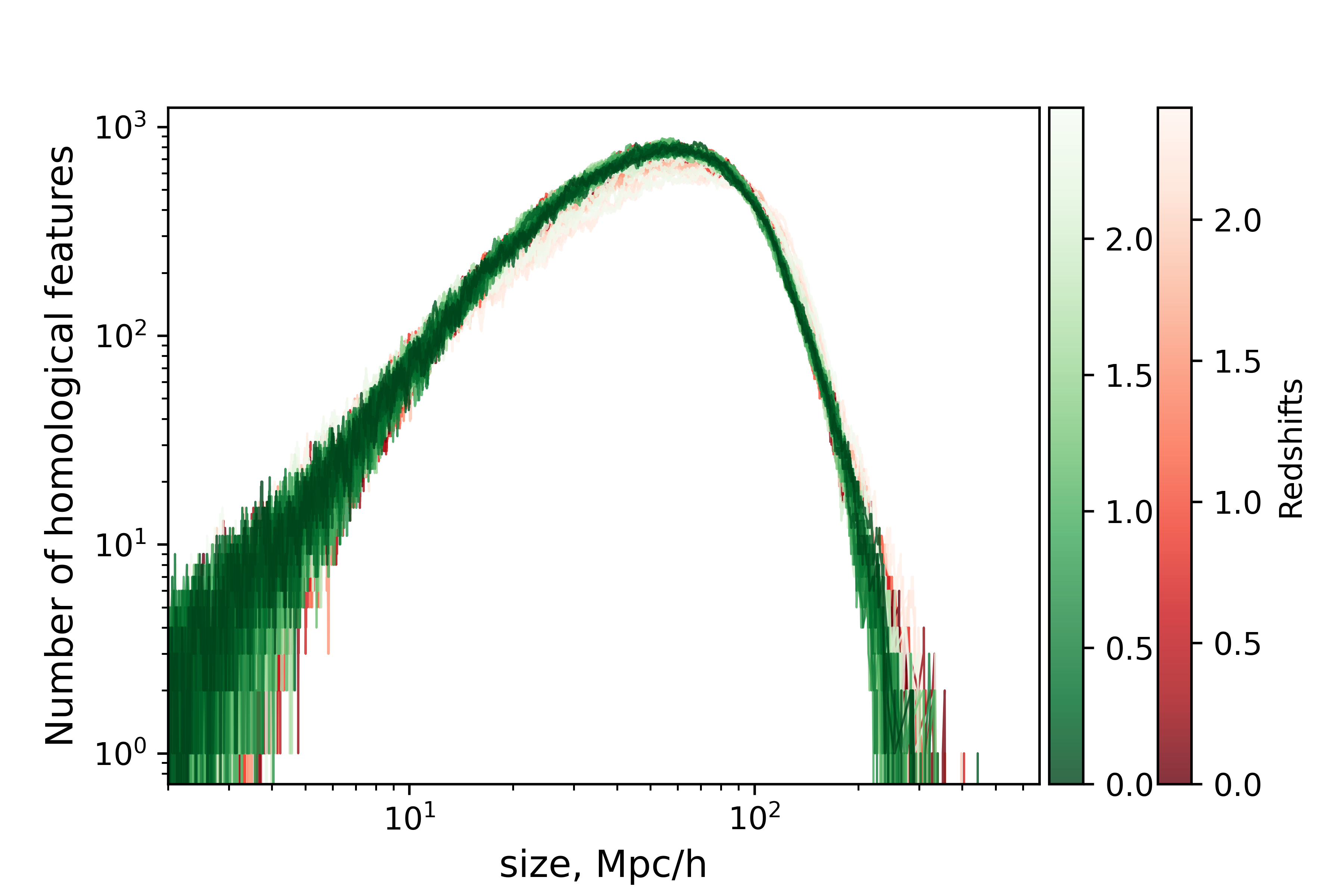}\\
\footnotesize Multidark vs. Bolshoi, $\ln(M/M_\odot) = 11.0\dots 11.5$ & \footnotesize Multidark vs. Bolshoi, $\ln(M/M_\odot) = 11.5\dots 12.0$\\
\includegraphics[width=0.48\textwidth]
{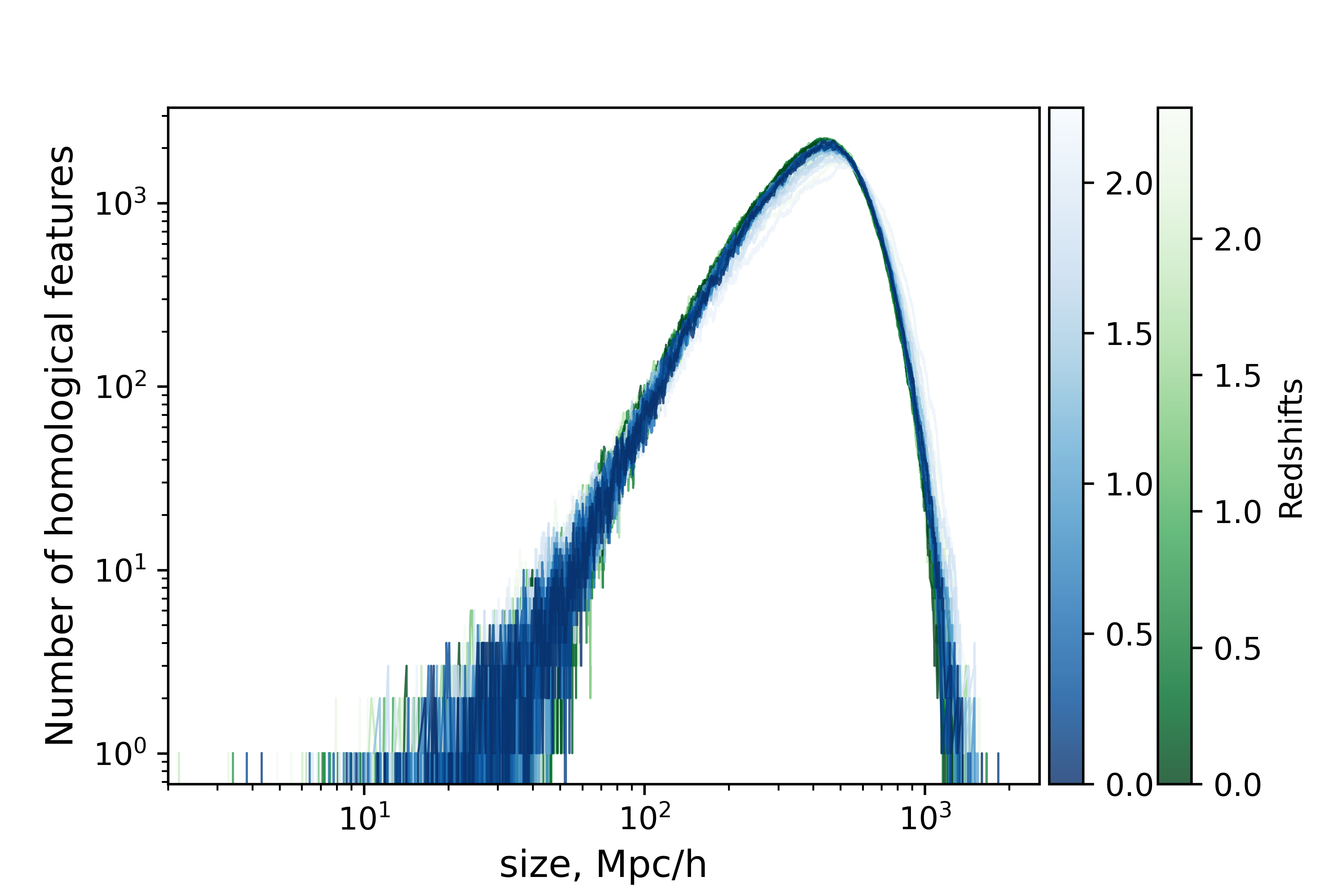} &
\includegraphics[width=0.48\textwidth]
{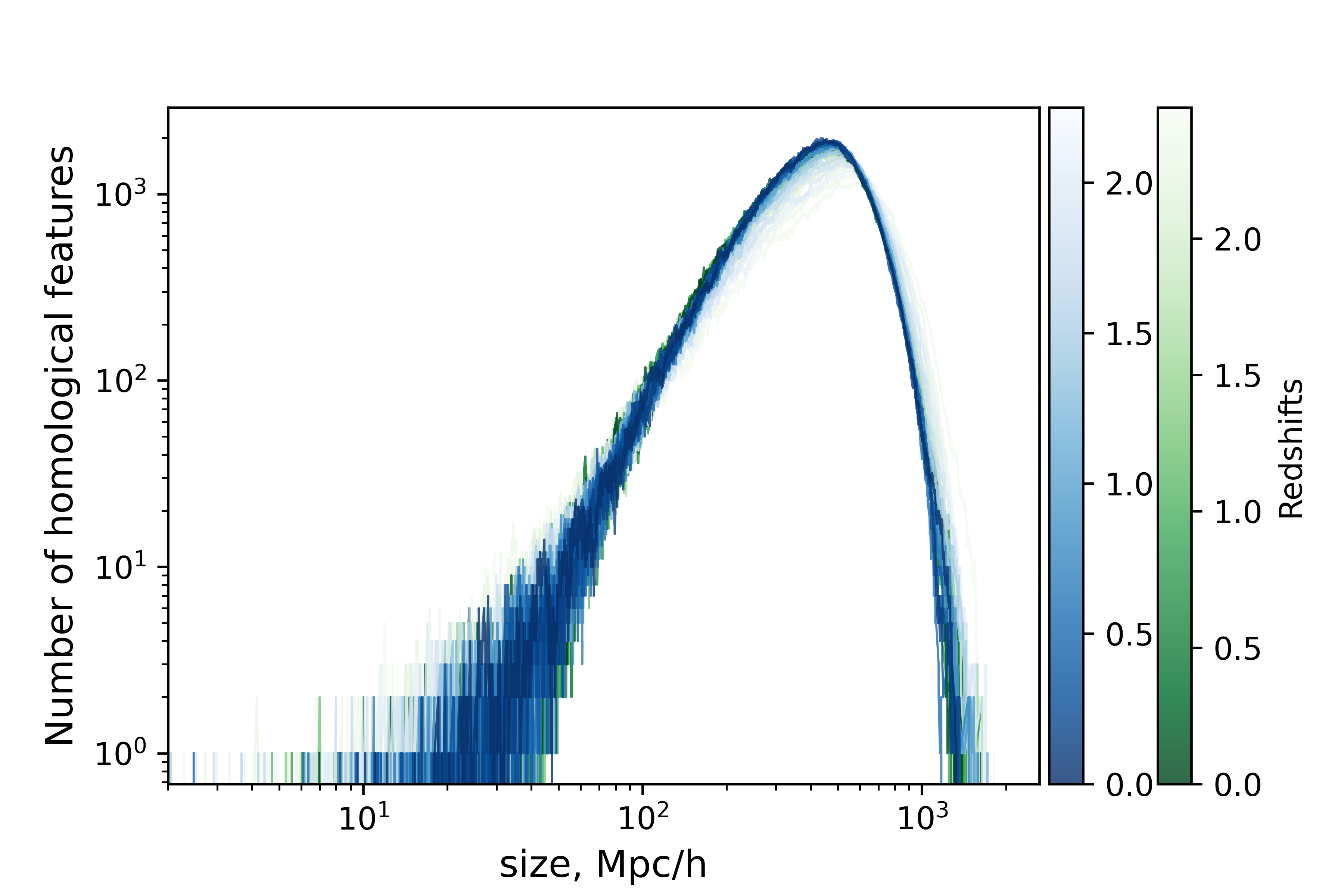}\\
\footnotesize HOWLS vs. Multidark, $\ln(M/M_\odot) = 12.0\dots 12.5$ & \footnotesize HOWLS vs. Multidark, $\ln(M/M_\odot) = 12.5\dots 13.0$
\end{tabular}
\caption{Comparison of Betti curves evolution for $2$-dimensional homology features. All graphs depict Bolshoi curves in red, Multidark in green, and HOWLS in blue.}
\label{bc_massranges}
\end{center}
\end{figure*}

From these figures, one could have concluded that the resolution of the simulation defines whether we see a specific evolution or not. However, if we recall on ``topological bias'' effect \cite{Bermejo22}, it seems to be more reasonable to explore Betti curves for subpopulations with a narrow mass range. 

Indeed, after this modification, the picture changes drastically\,--- curves become very stable, showing almost no evolution with cosmological time. In Fig.\ref{bc_massranges}, we show the evolution of Betti curves for $2$-dimensional topological features in the narrow mass ranges. In each range, only $50 000$ halos were randomly sampled to have an equal number of halos in each subpopulation, as the properties of the Betti curves heavily depend on the concentration of points in the poincloud. The comparison is made in the Bolshoi-Multdark pair (for a volume of $(250~\mathrm{Mpc/h})^3$) and Multidark-HOWLS pair (for a volume of $(750~\mathrm{Mpc/h})^3$). 

One can observe a new peculiar feature: Betti curves of different simulations are now much more similar, showing that subpopulations of the same mass range are not only almost not evolving, but they are practically the same for simulations of different resolutions. 
Once the simulation has a resolution high enough to be complete for a given mass range, the web generated by its halos attains a sort of "universality". This also means that the effect of cosmic variation plays a negligible role when comparing such subpopulations. The Cosmic webs of simulated halos thus appear to possess a new topological invariant, which is well preserved through the cosmological evolution at least for $z \leq 2.5$. 

\subsection{Approximation of Betti curves}

We have discovered that Betti curves of Cosmic webs are effectively approximated with an appropriately scaled log-normal probability distribution function
$$
f_{A,\mu,\sigma}(r) = \frac{A}{r} e^{-\frac{(\ln r - \mu)^2}{2 \sigma^2}}. 
$$
Here $r$ is the parameter of $\alpha$-complex, $A$ is the amplitude, while $\mu$ and $\sigma$ are called expected value (or mean) and standard deviation, respectively. Since $\alpha$-complex is based on halo positions\,--- $r$ should be measured in Mpc/h,--- we understand $\ln r$ in terms of quantity calculus, i.e. $\ln (r/1 \text{Mpc})$. Please note, if the random variable $X$ is log-normally distributed, then $Y = \ln(X)$ has a normal distribution, $\mu$, and $\sigma$ are mean value and standard deviation of $Y$, not $X$.

 Interestingly, log-normal distribution was already proposed for related subjects in \cite{Giri21}. However, in their work, the authors approximate the Betti function of continuous density perturbation fields. It is also well known that cosmic voids have a log-normal distribution of their radii. Therefore, the choice for the Betti curve approximation as a scaled log-normal distribution seems to be quite reasonable.

For approximation, we used \texttt{curve\_fit} from the \texttt{scipy.optimize} Python package. To get the initial point for minimization, we note that
$$
f(r) = A e^{\mu-\sigma^2/2} e^{-\frac{(\ln r - (\mu-\sigma^2))^2}{2\sigma^2}},
$$
which is a scaled pdf for normal distribution $g(u)$ for $u = \ln r$. Thus for each data point we change coordinates from $(x;y)$ to $(u=\ln x; y)$, perform normalization and calculate mean $\mu_g$ and variance $\sigma_g$ with the help of integration by \texttt{trapezoid} function from \texttt{scipy.integrate}
$$
A_g = \int\limits_0^\infty y(u) du;\quad
\mu_g = \frac{1}{A_g}\int\limits_0^\infty u y(u) du;\quad
\sigma_g = \frac{1}{A_g}\int\limits_0^\infty u^2 y(u) du - \mu_g^2.
$$
\begin{figure}[!htb]
\begin{center}
\begin{tabular}{lr}
\includegraphics[width=0.5\textwidth]
{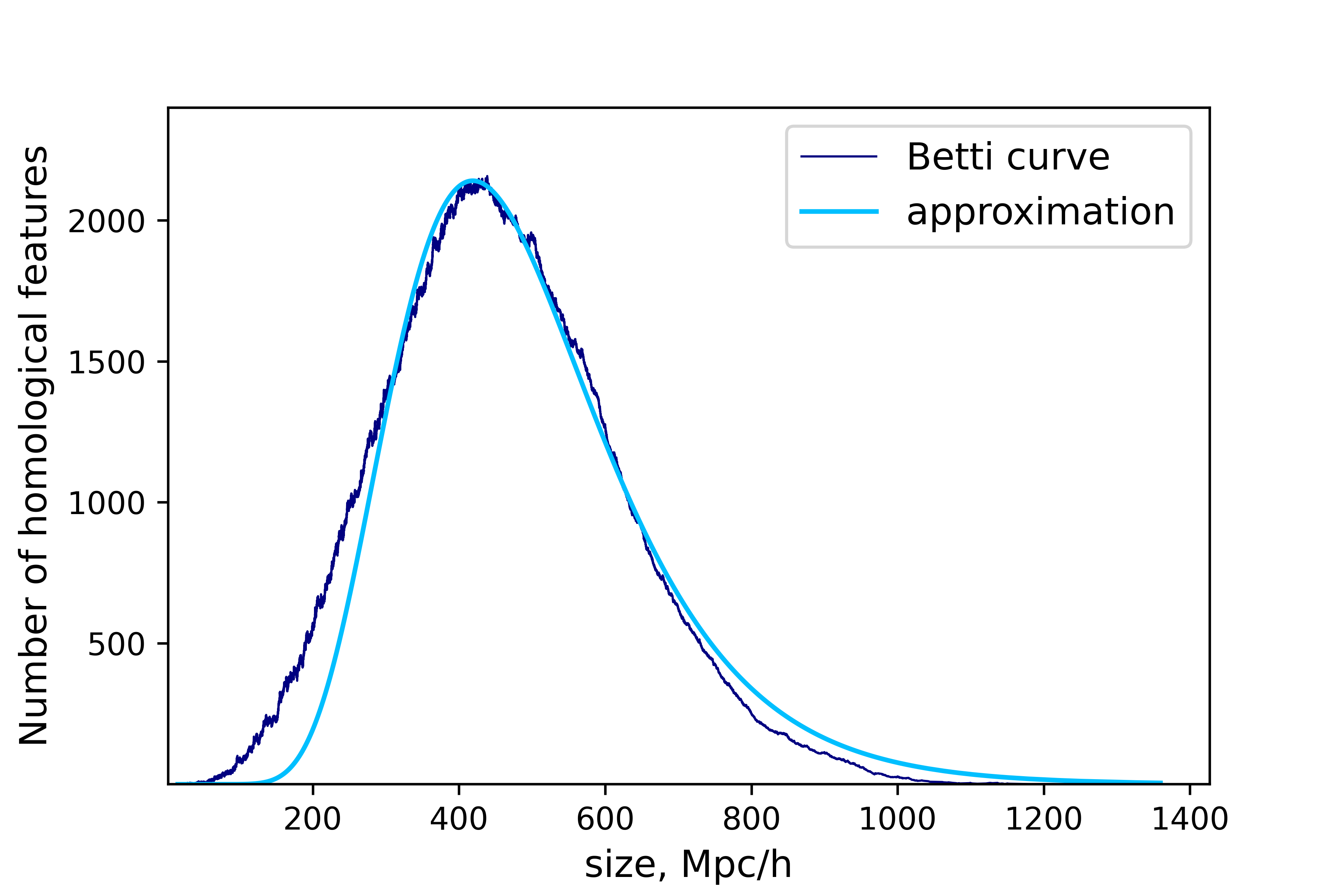} &
\includegraphics[width=0.5\textwidth]
{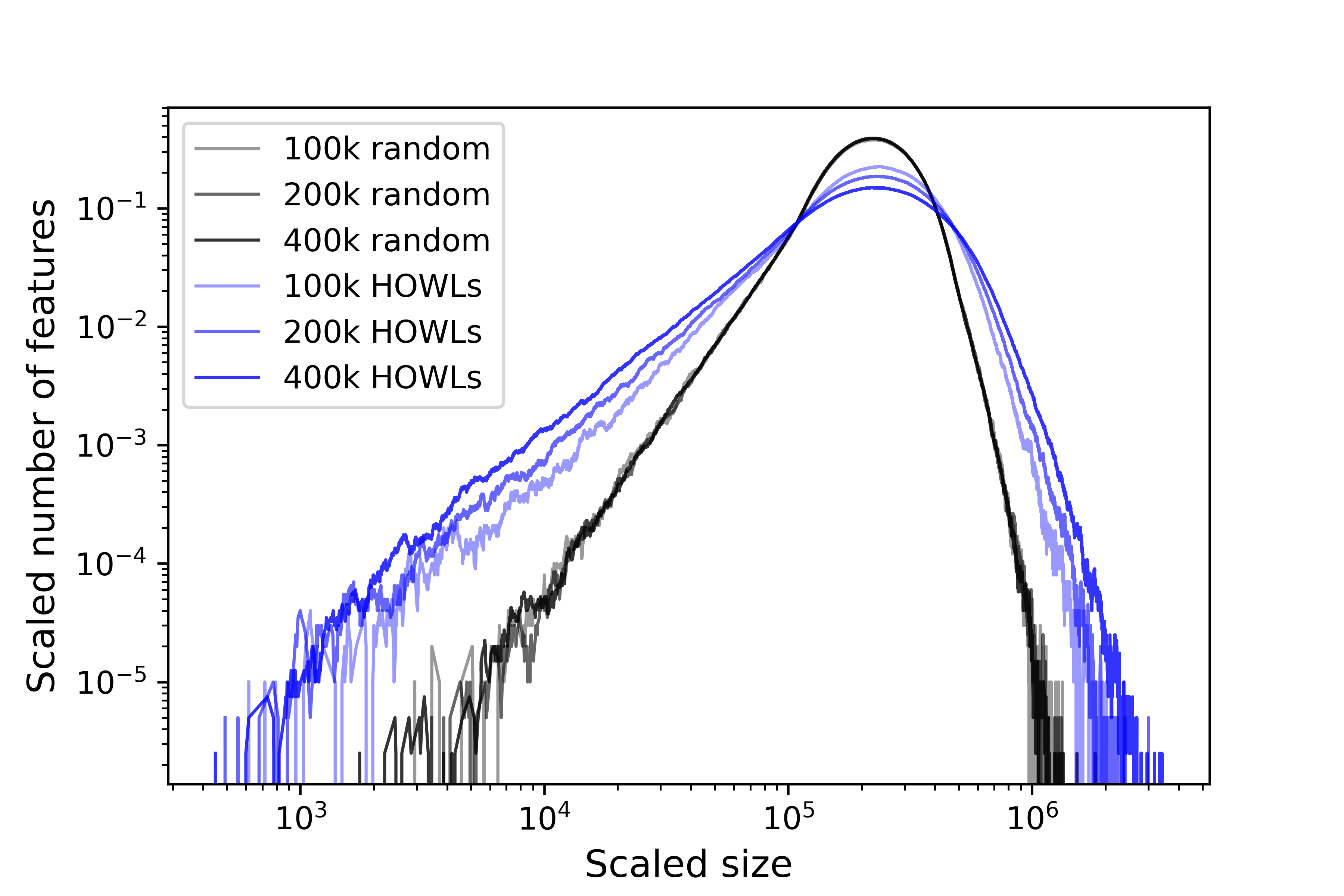}
\end{tabular}
\caption{Left: original Betti curve of dimension 2 homology features (voids) and its approximation with scaled log-normal function. Computed for halo population of Multidark universe at $z=0$, $R^2=0.96$. Right: Normalized Betti curves of random point cloud and HOWLS simulation's Cosmic web. The number of points are $100$, $200$, and $400$ thousand.}
\vspace{-0.8cm}
\label{betti_vs_approx}
\end{center}
\end{figure}

Initial values of parameters are then calculated as
$$
\sigma_I = \sigma_g;\qquad
\mu_I = \mu_g + \sigma_g^2;\qquad
A_I = \frac{A_g}{\sigma_I \sqrt{2 \pi}} e^{\mu_I-\sigma_I^2/2}.
$$
After initial values are calculated, they are passed together with raw data and model into \texttt{curve\_fit} from the \texttt{scipy.optimize} Python package that performs further refinement. For quality control, we calculated RMSE and $R^2$ (percent of explained variation). For the reader to assess the quality of the approximation, we present tab~\ref{tab:goodness_of_fit}, where we show the $R^2$ and tab~\ref{tab:parameters_mean} with values of mean, standard deviation, and range of for all approximated curves, averaged by $z$. An example of approximation is given in fig.\ref{betti_vs_approx}
\begin{table}[!htb]
\begin{tabular}{ |p{2cm}|p{3.5cm}|c|c|c|c|  }
\hline
\multirow{3}{2cm}{Model parameters}
& scale, Mpc/h & \multicolumn{4}{|c|}{$250$}\\
\cline{2-6}
& $\ln(M/M_\odot)$ range & \multicolumn{2}{|c|}{$11.0 \dots 11.5$} 
             & \multicolumn{2}{|c|}{$11.5 \dots 12.0$}\\
\cline{2-6}
& homology & $1$ & $2$ & $1$ & $2$\\
\hline
\multirow{3}{2cm}{Goodness of fit}
& $\mathbb{E}[R^2]$ & $0.98$
                    & $0.93$
                    & $0.97$
                    & $0.9$\\
\cline{2-6}  
& $\text{SD}[R^2]$ & $0.01$
                   & $0.02$
                   & $0.02$
                   & $0.02$\\
\cline{2-6}
& $\text{Range}[R^2]$ & $0.95 \dots 0.99$
                      & $0.86 \dots 0.96$
                      & $0.90 \dots 0.99$
                      & $0.82 \dots 0.94$\\
\hline
\multirow{3}{2cm}{Model parameters}
& scale, Mpc/h & \multicolumn{4}{|c|}{$750$}\\
\cline{2-6}
& $\ln(M/M_\odot)$ range & \multicolumn{2}{|c|}{$12.0 \dots 12.5$} 
             & \multicolumn{2}{|c|}{$12.5 \dots 13.0$}\\
\cline{2-6}
& homology & $1$ & $2$ & $1$ & $2$\\
\hline

\multirow{3}{2cm}{Goodness of fit}
& $\mathbb{E}[R^2]$ & $0.98$
                    & $0.95$
                    & $0.98$
                    & $0.94$\\
\cline{2-6}  
& $\text{SD}[R^2]$ & $0.003$
                   & $0.01$
                   & $0.01$
                   & $0.01$\\
\cline{2-6}
& $\text{Range}[R^2]$ & $0.97 \dots 0.99$
                      & $0.92 \dots 0.97$
                      & $0.96 \dots 0.99$
                      & $0.91 \dots 0.96$\\
\hline
\end{tabular}
\caption{Values of $R^2$ (percent of explained variation) for approximations of Betti curves.}
\label{tab:goodness_of_fit}
\end{table}

\begin{figure*}[!htb]
\begin{center}
\begin{tabular}{cc}
\includegraphics[width=0.48\textwidth]
{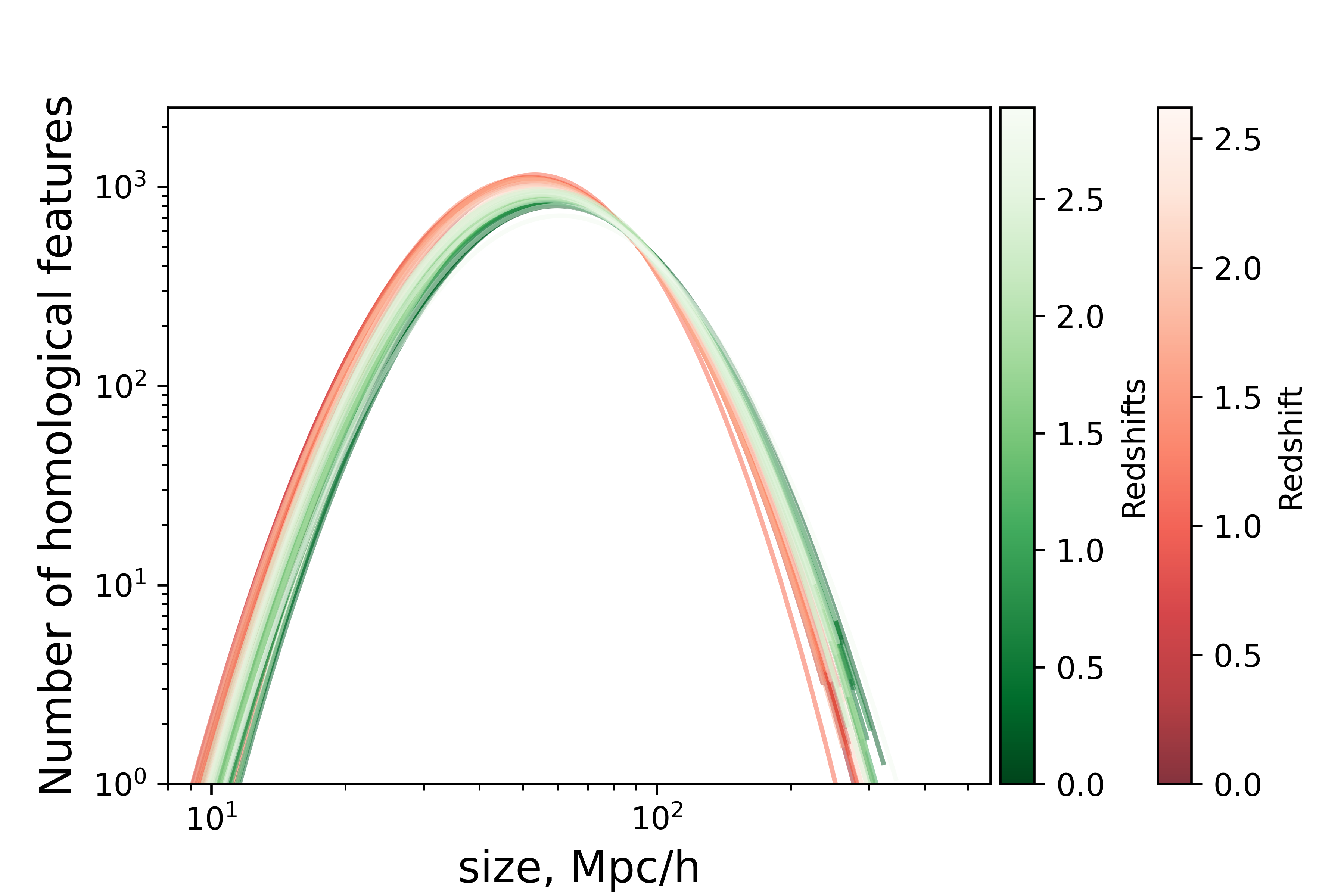}&
\includegraphics[width=0.48\textwidth]
{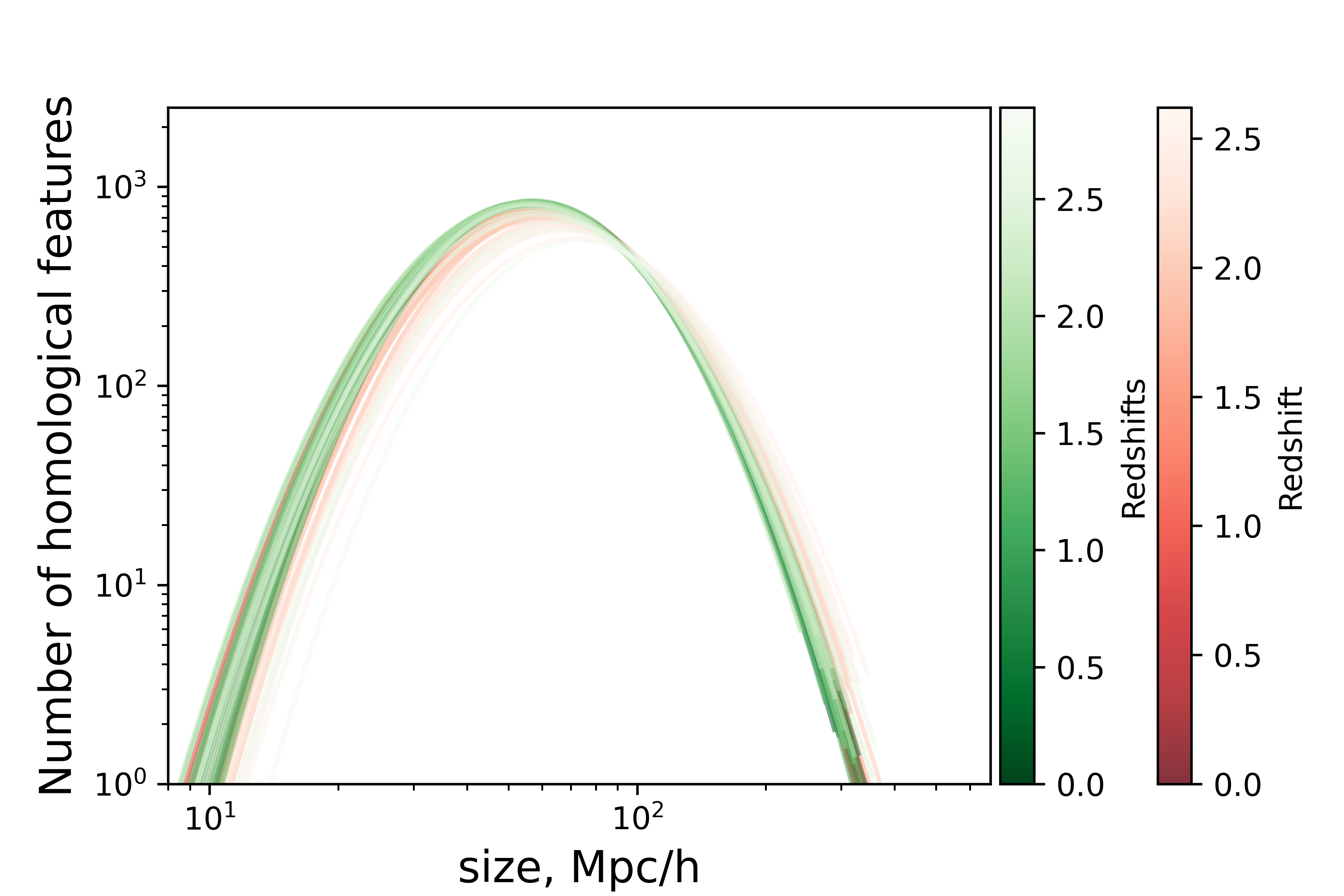}\\
\footnotesize Multidark vs. Bolshoi, $\ln(M/M_\odot) = 11.0\dots 11.5$ & \footnotesize Multidark vs. Bolshoi, $\ln(M/M_\odot) = 11.5\dots 12.0$\\
\includegraphics[width=0.48\textwidth]
{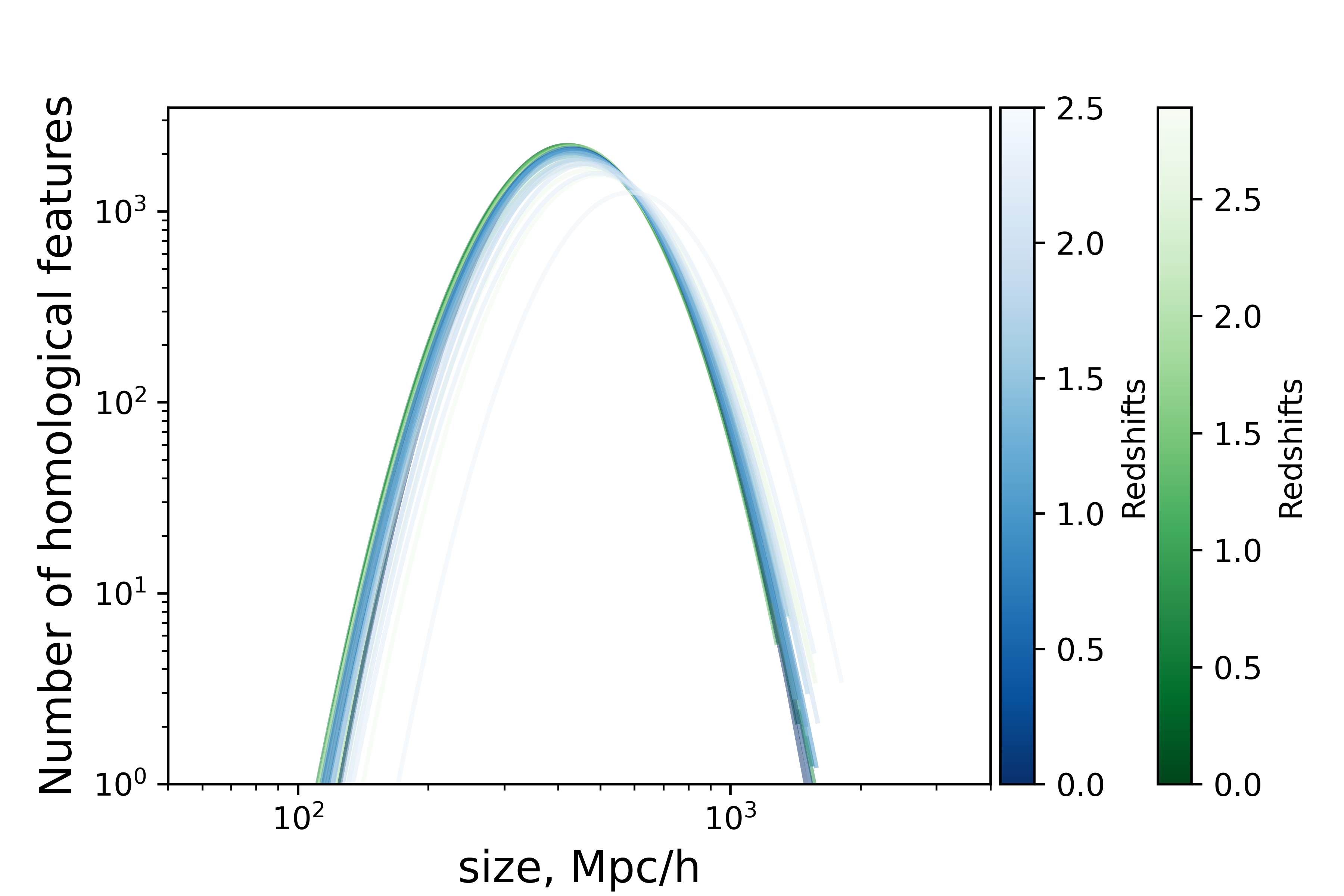}&
\includegraphics[width=0.48\textwidth]
{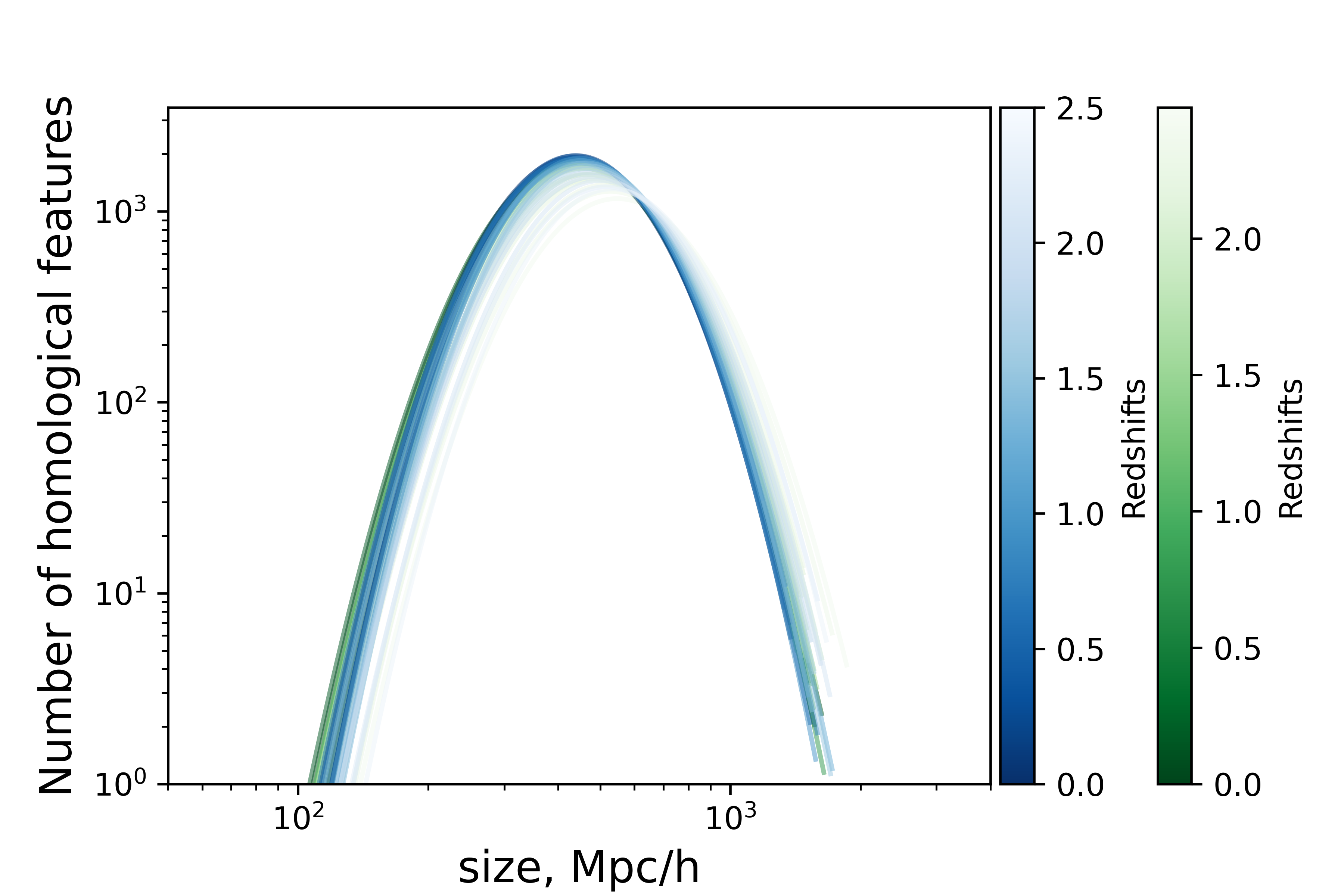}\\
\footnotesize HOWLS vs. Multidark, $\ln(M/M_\odot) = 12.0\dots 12.5$ & \footnotesize HOWLS vs. Multidark, $\ln(M/M_\odot) = 12.5\dots 13.0$
\end{tabular}
\caption{Comparison of approximated Betti curves evolution for $2$-dimensional homology features. All graphs depict Bolshoi curves in red, Multidark in green, and HOWLS in blue.}
\label{bc_approx}
\end{center}
\end{figure*}

\begin{figure*}
\begin{center}
\begin{minipage}{0.7\textwidth}
    \includegraphics[width=\textwidth]{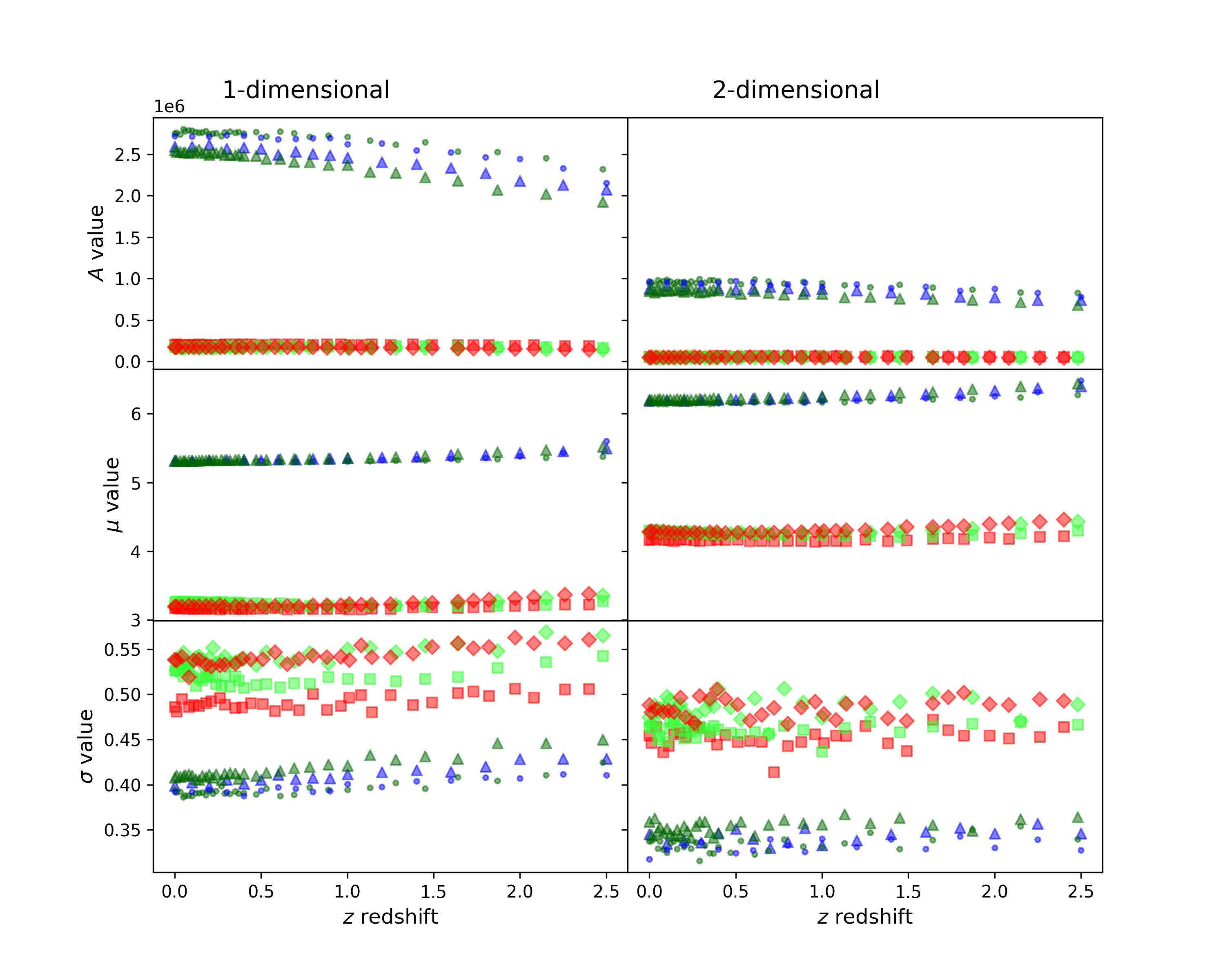}
\end{minipage}
\fbox{\begin{minipage}{0.3\textwidth}
\footnotesize
\begin{tabular}{rl}
\mysquare{red} & Bolshoi, 250 Mpc/h,\\
& $\ln(M/M_\odot) = 11.0\dots11.5$,\\
\mysquare{green} & Multidark, 250 Mpc/h,\\
& $\ln(M/M_\odot) = 11.0\dots11.5$,\\
\mydiamond{red} & Bolshoi, 250 Mpc/h,\\
& $\ln(M/M_\odot) = 11.5\dots12.0$,\\
\mydiamond{green} & Multidark, 250 Mpc/h,\\
& $\ln(M/M_\odot) = 11.5\dots12.0$,\\
\mycircle{blue} & HOWLS, 750 Mpc/h,\\
& $\ln(M/M_\odot) = 12.0\dots12.5$,\\
\mycircle{teal} & Multidark, 750 Mpc/h,\\
& $\ln(M/M_\odot) = 12.0\dots12.5$,\\
\mytriangle{blue} & HOWLS, 750 Mpc/h,\\
& $\ln(M/M_\odot) = 12.5\dots13.0$,\\
\mytriangle{teal} & Multidark, 750 Mpc/h,\\
& $\ln(M/M_\odot) = 12.5\dots13.0$. \\
\end{tabular}
\end{minipage}}
\caption{Evolution of approximation parameters of Betti curves.}
\label{params}
\end{center}
\end{figure*}

In fig.\ref{bc_approx}, one can find the approximations of Betti curves, pictured in fig. \ref{bc_massranges}. Curves look very similar (as expected). To be more qualitative about their evolution, we plot parameters of the Betti curves approximations on the redshift in fig.\ref{params}.

What can one infer from this picture? We see that all three parameters of approximation exhibit more or less constant behavior, with mean value $\mu$ being the most stable and dispersion $\sigma$ being the most noisy. The amplitude parameter $A$ shows a monotonic evolution, increasing with cosmic time.
As we have already argued, such noisy behavior can be explained by short-living homology features. Naturally, the presence of these features leads to broadening the Betti curves. This is why the dispersion parameter absorbs impact from them. Increasing amplitude and a slight decrease of dispersion mean that during the evolution, mass ranges more strictly choose "their" scale as Betti curves become narrower and higher. Nevertheless, the deviation measured during the cosmological evolution never exceeds 15$\%$ from the mean value, as can be seen in the tab. \ref{tab:parameters_mean}. We can also see that features of lower dimensions occupy a lower scale. Betti curves of the 1-dimensional feature have smaller mean values and are narrower ($A$ to $\sigma$ relation) than the 2-dimensional ones.
\begin{table}
\begin{tabular}{ |p{2cm}|p{3.2cm}|c|c|c|c|  }
\hline
\multirow{3}{2cm}{Model parameters}
& scale, Mpc/h & \multicolumn{4}{|c|}{$250$}\\
\cline{2-6}
& $\ln(M/M_\odot)$ range & \multicolumn{2}{|c|}{$11.0 \dots 11.5$} 
             & \multicolumn{2}{|c|}{$11.5 \dots 12.0$}\\
\cline{2-6}
& homology & $1$ & $2$ & $1$ & $2$\\
\hline
\multirow{6}{2cm}{Fit parameters}
& $\mathbb{E}[A]$, Mpc/h
& $1.9 \cdot 10^5\, 
  {{\color{blue}-12\%} \above 0pt {\color{red}+10\%}}$
& $5.9 \cdot 10^4\,
{{\color{blue}-10\%} \above 0pt {\color{red}+13\%}}$
& $1.7 \cdot 10^5\,
  {{\color{blue}-15\%} \above 0pt {\color{red}+5\%}}$
& $5.1 \cdot 10^4\,
  {{\color{blue}-11\%} \above 0pt {\color{red}+6\%}}$\\
\cline{2-6}
& $\text{SD}[A]$, Mpc/h & $1.3 \cdot 10^4$
                      & $3.6 \cdot 10^3$
                      & $7.7 \cdot 10^3$
                      & $1.8 \cdot 10^3$\\
\cline{2-6}
& $\mathbb{E}[\mu]$ 
& $3.21\,
  {{\color{blue}-2\%} \above 0pt {\color{red}+2\%}}$
& $4.22\,
  {{\color{blue}-2\%} \above 0pt {\color{red}+2\%}}$
& $3.22\,
  {{\color{blue}-1\%} \above 0pt {\color{red}+5\%}}$
& $4.29\,
  {{\color{blue}-1\%} \above 0pt {\color{red}+4\%}}$
\\
\cline{2-6}
& $\text{SD}[\mu]$ & $0.04$
                   & $0.05$
                   & $0.05$
                   & $0.05$\\
\cline{2-6}
& $\mathbb{E}[\sigma]$
& $0.506\,
  {{\color{blue}-5\%} \above 0pt {\color{red}+7\%}}$
& $0.456\,
  {{\color{blue}-9\%} \above 0pt {\color{red}+4\%}}$
& $0.542\,
  {{\color{blue}-4\%} \above 0pt {\color{red}+5\%}}$
& $0.485\,
  {{\color{blue}-6\%} \above 0pt {\color{red}+5\%}}$
\\
\cline{2-6}
& $\text{SD}[\sigma]$ & $0.016$
                      & $0.01$
                      & $0.01$
                      & $0.011$\\
\hline
\multirow{3}{2cm}{Model parameters}
& scale, Mpc/h & \multicolumn{4}{|c|}{$750$}\\
\cline{2-6}
& $\ln(M/M_\odot)$ range & \multicolumn{2}{|c|}{$12.0 \dots 12.5$} 
             & \multicolumn{2}{|c|}{$12.5 \dots 13.0$}\\
\cline{2-6}
& homology & $1$ & $2$ & $1$ & $2$\\
\hline
\multirow{6}{2cm}{Fit parameters}
& $\mathbb{E}[A]$, Mpc/h
& $2.7 \cdot 10^6\,
  {{\color{blue}-19\%} \above 0pt {\color{red}+5\%}}$
& $9.3 \cdot 10^5\,
  {{\color{blue}-17\%} \above 0pt {\color{red}+6\%}}$
& $2.4 \cdot 10^6\,
  {{\color{blue}-21\%} \above 0pt {\color{red}+8\%}}$
& $8.3 \cdot 10^5\,
  {{\color{blue}-18\%} \above 0pt {\color{red}+9\%}}$\\
\cline{2-6}
& $\text{SD}[A]$, Mpc/h & $1.4 \cdot 10^5$
                      & $4.7 \cdot 10^4$
                      & $1.6 \cdot 10^5$
                      & $5.1 \cdot 10^4$\\
\cline{2-6}
& $\mathbb{E}[\mu]$ 
& $5.32\,
  {{\color{blue}-1\%} \above 0pt {\color{red}+5\%}}$
& $6.18\,
  {{\color{blue}-1\%} \above 0pt {\color{red}+5\%}}$
& $5.35\,
  {{\color{blue}-1\%} \above 0pt {\color{red}+3\%}}$
& $6.25\,  
  {{\color{blue}-1\%} \above 0pt {\color{red}+3\%}}$\\
\cline{2-6}
& $\text{SD}[\mu]$ & $0.05$
                   & $0.05$
                   & $0.05$
                   & $0.06$\\
\cline{2-6}
& $\mathbb{E}[\sigma]$
& $0.396\,
  {{\color{blue}-2\%} \above 0pt {\color{red}+7\%}}$
& $0.333\,
  {{\color{blue}-5\%} \above 0pt {\color{red}+6\%}}$
& $0.415\,
  {{\color{blue}-4\%} \above 0pt {\color{red}+8\%}}$
& $0.35\,
  {{\color{blue}-6\%} \above 0pt {\color{red}+5\%}}$\\
\cline{2-6}
& $\text{SD}[\sigma]$ & $0.008$
                      & $0.008$
                      & $0.012$
                      & $0.009$\\
\hline
\end{tabular}
\caption{Mean, standard deviation, and range of approximating functions' parameters through the cosmological evolution of Betti curves.
Notation $x\,{{\color{blue}-y\%} \above 0pt {\color{red}+z\%}}$ for $\mathbb{E}[X]$ means $\mathbb{E}[X] = x$, while $\min(X) = x - x\cdot\frac{y}{100}$ and $\max(X) = x + x\cdot\frac{z}{100}$.
One can see that the maximal possible deviation for $\mu$ and $\sigma$ from their respective mean values is within several percent for all redshifts $z \leq 2.5$.
}
\label{tab:parameters_mean}
\end{table}
Finally, we can observe that the parameters' values strongly depend on the simulation box size. This effect follows from the well-known fact that Betti curves are highly dependent on the concentration of particles in a point cloud. 

But the general feature to be pointed out here is that Betti curves of the Cosmic webs are sufficiently stable under cosmic evolution. We can see the mass ``splitting'' in the behavior of $\mu$, which was earlier noticed by other authors \cite{Bermejo22} and dubbed as the ``topological bias'' - the subpopulation of halos of each mass range forms the structures of their own scale. The evolution of approximation parameters inside a given mass range can be explained by changes in mass function even within this already narrow mass range. 

\begin{figure*}[!htb]
\begin{center}
\includegraphics[width=\textwidth]
{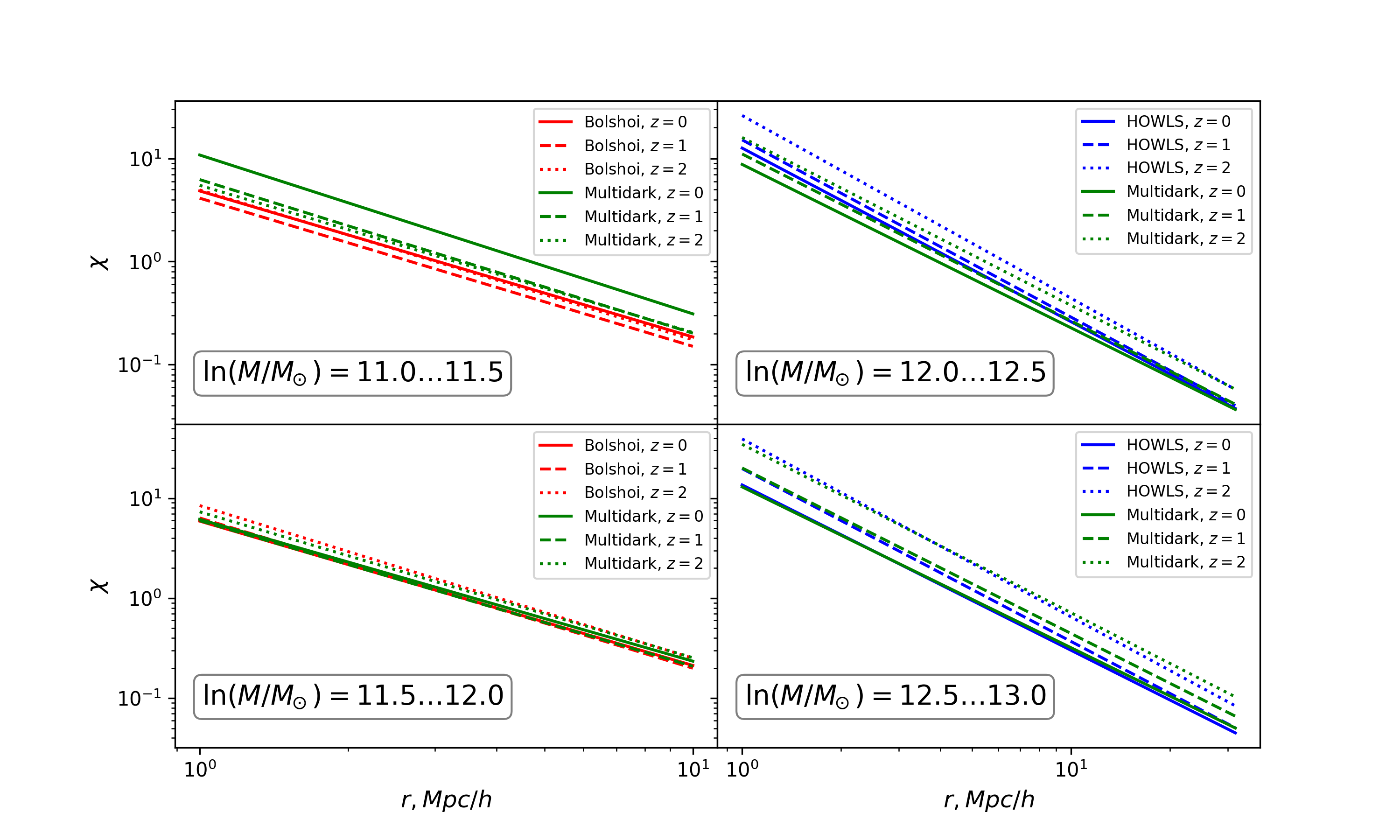}
\caption{
Evolution of two-point correlation function for different simulations and mass ranges.
Evaluation performed with the Landy-Szalay estimator \cite{Landy93}, the number of random points taken exceeds $50$ times the number of data points.
}
\label{fig:corrf}
\end{center}
\end{figure*}

\subsection{Comparison with random point cloud}

We want to emphasize that Betti curves of Cosmic web are radically different from those of random point clouds, not only in their shape and form but also in their behavior in scaling. Let us compare them in fig. \ref{betti_vs_approx}.

We generated three populations with 100, 200, and 400 thousand points randomly positioned in the cubes, identical in size to cubes of the HOWLS simulation. On the graph, we compare the 1-dimensional Betti curves of random point clouds (black lines) with those of the HOWLS halo population, with the same number of halos (blue lines) randomly sampled from $z=0$ HOWLS universe. The x-axis on this plot (persistence radii) was normalized by the number of points in the poincloud raised to the power 2/3, and the y-axis - by the number of points. The explanation for such an effect of scaling for Betti curve of a random point cloud is given in \cite{Gluzberg2023}. The shape and position of the maximum of a random point cloud Betti curve remain the same if such normalization is applied. However, as can be seen from fig.\ref{betti_vs_approx}, this normalization will not arrange the Betti curves of Cosmic web. The scaling of Betti curves of Cosmic webs requires exploration in a separate study. 

\section{Conclusions and discussion}
\label{sec:conlusions}
In this study, we have explored the Betti curves of the Cosmic web of different cosmological simulations. To take into account the so-called "topological bias," we considered several subpopulations of specific mass ranges of halos. We have discovered that mass range and the distribution of mass inside this mass range completely determine this population's Betti curves. On the contrary, Betti curves have no strong dependence on the redshift. We can thus say that Betti curves do not evolve with cosmic time inside a given mass range. In this sense, our analysis has unveiled the existence of a new topological invariant in the description of the Cosmic web. Moreover, such property of Betti curves becomes an advantage when comparing them to the correlation functions of halo, see Fig.\ref{fig:corrf}. One can note, that (linear approximation of) correlation functions evolution with redshift is not always monotonic, for example, for Bolshoi and Multidark simulations in 11.5-12.0 $\log M$ mass range. On the contrary, the Betti curves fit parameters behavior with $z$ evidence clearly that their evolution with cosmic time is weak (and still monotonic), as can be seen from Fig.\ref{params}, they are nearly constant in redshift.

Finally, as it was formulated in \cite{Wilding21}, we can confirm that "the structure on small scales at earlier times resembles that of the structure on large scales at later times" - as the masses of the halos grow with cosmic time, the same mass range will take different places in the hierarchy of large scale structures. This conclusion can be useful when studying the Cosmic web with topological data analysis tools. The coupling between masses of population and topological features of its web is crucial, and understanding it will help one avoid biased inferences in the analysis. We can also claim that the Betti curves of the Cosmic web at the considered scale are resistant to the cosmic variation: all the simulations had different initial (seed) conditions, yet the resemblance of their Betti curves is rather spectacular. 

We have found that higher-dimension structures (voids) exist on larger scales than lower-dimensional ones (cycles and tunnels of filaments). Betti curves of both of them can be approximated with a scaled log-normal function with three parameters (amplitude, mean, and dispersion of a curve). The Betti curves of lower dimension seem to be sharper: they have higher amplitudes and lower mean values and dispersions. 

There is a light sign of evolution through cosmological time in parameters of approximating function for both dimensions: the amplitude tends to become higher and mean values - to become lower. The dissimilarities, which are surely present, can be further analyzed. In particular, it will be interesting to find dependence on the halo mass function inside a narrow mass range on the Betti curve.

The comparison of Cosmic web Betti curves and those of random point clouds reveals that they have different shapes and different behaviors with scaling regarding the concentration of points in the point cloud. This, as well as developing quantitative methods of Betti curves usage in discriminating cosmological parameters, will be the subjects of future works.    

\section{Acknowledgements}
We would like to thank the Armed Forces of Ukraine for providing security to perform this work. This work was supported by the National Research Foundation of Ukraine under Project No. 2020.02/0073.

F.\,V. has been supported by Fondazione Cariplo and Fondazione CDP, through grant n° Rif: 2022-2088 CUP J33C22004310003 for "BREAKTHRU" project.

V.\,T. wants to express his gratitude to Dr. Sergiy Maksymenko (Institute of Mathematics, NAS of Ukraine) for his valuable comments and feedback on the Appendix~\ref{sec:appendixa} that greatly improved the clarity and quality of the material.

\section*{Appendix A. A more rigorous exposition of Betti curves}
\label{sec:appendixa}

In this section, we provide a more rigorous exposition of the method used.
It is intended to provide more insights for readers who are not familiar with homology theory but want a deeper understanding of the methods employed. We will suppose that Reader is already familiar with some basic definitions of topology and algebra like ``topological space'' and ``abelian group''.
For those Readers who want even more details, we recommend books~\cite{maunder,nash,chasal,hatcher}.

\subsection{Simplicial Complex: a Bridge between Topology and Algebra}

The first step in the analysis provided is the construction of a special mathematical object\,--- a simplicial complex\,--- from the dataset given.
A simplicial complex is so special because it bridges topology and algebra.
Let's first define a simplex as a geometrical object.

\begin{dfn} 
A set of points $\{a_0,\dots,a_n\}$ in some Euclidean space $\mathbb{R}^m$ is said to be \textbf{independent} if vectors $a_1 - a_0$, $a_2 - a_0$,...,$a_n - a_0$ are linearly independent.
\end{dfn}

\begin{thm}
Any subset of a linearly independent set of points is linearly independent.
\textnormal{$\blacktriangleleft$ Obvious if $a_0$ is in the subset. 
If not take any $a_k$ and consider vectors $a_i - a_k$ together with $a_k - a_0$. 
This set is linearly independent due to $(a_i - a_k) + (a_k - a_0) = a_i - a_0$, thus a set of $a_i - a_k$ only is linearly independent as well. $\blacksquare$}
\end{thm}

\begin{dfn} 
A \textbf{geometric $n$-simplex} $\sigma^n$ is a set of points
$$
\sigma^n = \left\{\left.
\sum_{i=0}^n \lambda_i a_i\right|
\lambda_i \geq 0,\quad \sum_{i=0}^n \lambda_i = 1
\right\}
$$
where $a_i$ are independent points in some Euclidean space $\mathbb{R}^m$; $\sigma^n$ is given the subspace topology.
The subspace of $\sigma^n$ of the points with  $\lambda_{k_0}=0,\dots,\lambda_{k_p}=0$ for a set of indices $\{k_0,\dots,k_p\}$ is called a \textbf{face of $\sigma^n$}.
The face is \textbf{proper} if it's not empty or the whole $\sigma^n$.
\end{dfn} 

\begin{dfn} 
A \textbf{geometric simplicial complex $K$} is a finite set of simplices all contained in $\mathbb{R}^m$ and satisfying:
\begin{itemize}
    \item if $\sigma^n$ is a simplex of $K$ and $\tau^p$ is a face of $\sigma^n$, then $\tau^p$ is in $K$;
    \item if $\sigma^n$ and $\tau^p$ are simplices in $K$, then $\sigma^n \cap \tau^p$ is either empty or a common face of $\sigma^n$ and $\tau^p$.
\end{itemize}
The \textbf{dimension} of $K$ is the maximal dimension of its simplices.
\end{dfn}

\begin{dfn}
A union of all geometric simplices of a geometric simplicial complex $K$ equipped with a subset topology (with respect to $\mathbb{R}^m$) is called a \textbf{polyhedron $|K|$}.
\end{dfn}

But what if we consider a simplicial complex as a purely combinatorial object that only defines relations between its vertices without referring to geometry?

\begin{dfn} 
\label{def_abstract_simplex}
An \textbf{abstract simplicial complex $\mathcal{K}$} is a finite set of elements $a_0,\dots,a_N$ called \textbf{(abstract) vertices}, together with a collection of subsets of the form $(a_{i_0},\dots,a_{i_n})$ called \textbf{(abstract) simplices} closed under the operation of taking a subset (subset of a simplex is itself a simplex). 
A simplex $(a_{i_0},\dots,a_{i_n})$ contains $n+1$ points, while $n$ is called the \textbf{dimension of the simplex}. 
The \textbf{dimension of $\mathcal{K}$} is the maximum of the dimensions of its simplices.
\end{dfn}

\begin{dfn} 
Let $K$ be a geometric simplicial complex and $\mathcal{K}$ be an abstract simplicial complex such that there exists a bijection between their vertices and a subset of vertices being simplex in $\mathcal{K}$ if and only if they correspond to the vertices of some simplex in $K$. $\mathcal{K}$ is called an \textbf{abstraction} of $K$ and $K$ is called a \textbf{realization} of $\mathcal{K}$. 
\end{dfn}

Obviously, any geometric simplicial complex can be abstracted, but surprisingly we can always do the other way around with Theorem~\ref{thm_simplex_real}.
The realization is unique in some sense, thus abstract and geometrical simplices are firmly connected.

\begin{dfn}
Given geometric simplicial complexes $K$ and $L$, a \textbf{simplicial map} is a function $f: K \to L$ with the following properties:
\begin{itemize}
    \item if $a$ is a vertex of $K$, then $f(a)$ is a vertex of $L$;
    \item if $a_0$,...,$a_n$ are vertices of a simplex $\sigma^n$ of $K$, then $f(a_0)$,...,$f(a_n)$ span a simplex in $L$ (note: repeats possible) and $f(\sum_{i=0}^n \lambda_i a_i) = \sum_{i=0}^n \lambda_i f(a_i)$ (``linear'' on each simplex).
\end{itemize}
\end{dfn}

\begin{thm} 
\label{thm_simplex_real}
An $n$-dimensional abstract simplex $\mathcal{K}$ has a realization in $\mathbb{R}^{2n+1}$.
Moreover, let $K_1$ and $K_2$ be realisations of $\mathcal{K}$, then there exists a homeomorphism being a simplicial map $f : |K_1| \to |K_2|$.
\textnormal{$\blacktriangleleft$ see~\cite{maunder} $\blacksquare$}
\end{thm}

The key point is that the geometric information is retained within the abstraction of a geometric simplicial complex. If any information were lost during the abstraction, we should have been able to construct at least two non-homeomorphic realizations of $\mathcal{K}$ that fill in differently the missing piece of information. However, the Theorem~\ref{thm_simplex_real} forbids that, thus abstraction retains all the topological information. This connection paves the way for a transition between geometry and algebra.

\subsection{Doing Algebra on Simplicial Complexes}

Now we know how to create simplicial complexes from the data and we can study the topology of the appropriate polyhedrons.
Ironically, to study the topology of the simplicial complexes we need to transfer to algebra: the ultimate goal of this appendix\,--- the definition of Betti numbers\,--- is achieved in the realm of algebra.

\begin{dfn} 
\label{def_ordered_simplex}
An \textbf{oriented simplex $\sigma^n$} is an abstract simplex (Definition~\ref{def_abstract_simplex}) with orientation chosen. 
That is all the vertices of $\sigma^n$ are arbitrarily ordered, say $[v_0,\dots,v_p]$, and this order is given sign $+$. 
For any different ordering of the vertices, the sign is $+$ if it can be obtained from the chosen ordering by an even number of swaps of two vertices at a time, otherwise, it is $-$. 
Obviously, an \textbf{oriented abstract simplicial complex $\mathcal{K}$} is constructed from oriented simplices.
\end{dfn}

\begin{dfn} 
\label{def_chain}
The \textbf{$p$-chain} of $\mathcal{K}$, $C_p(\mathcal{K})$, is a free finitely generated abelian group (formally $\mathbb{Z}$-module), generated by oriented $p$-simplices of $\mathcal{K}$
$$
C_p(\mathcal{K}) = \left\{
\left.
\sum_{i=1}^{l_p} f_i \sigma_i^p
\right|
\forall i: f_i \in \mathbb{Z},~ \sigma_i^p \in \mathcal{K},~\sigma_i^p+\underbrace{(-\sigma_i^p)}_{\lefteqn{\scriptstyle\text{orientation}}}=0,~ 0\,\sigma_i^p = 0
\right\},
$$
where $l_p$ is the number of $p$-simplices in $\mathcal{K}$ and group operation ``$+$'' is defined as
$$
c = \sum_{i=1}^{l_p} f_i \sigma_i^p,\quad k = \sum_{i=1}^{l_p} g_i \sigma_i^p, \quad c + k = \sum_{i=1}^{l_p} (f_i + g_i) \sigma_i^p.
$$
For $p$ larger than the dimension of $\mathcal{K}$ we define $C_p(\mathcal{K})=\{0\}$.
\end{dfn}

Definitions~\ref{def_ordered_simplex} and~\ref{def_chain} effectively transform a simplicial complex into an abelian group. 
But it turns out to be not enough to characterize the topology.
Betti numbers characterize ``holes'' in a manifold, so we need something to probe for the boundary\,--- the boundary operator.

\begin{dfn}
The \textbf{boundary operator $\partial_p$} is the map $\partial_p: C_p(\mathcal{K}) \to C_{p-1}(\mathcal{K})$ such that
\begin{itemize}
    \item basis, i.e. oriented simplices, are transformed as follows
    $$
    \partial_p \sigma^p = \partial_p \underbrace{[v_0,\dots,v_p]}_{\text{ordered vertices}} = \sum_{i=0}^p (-1)^i [v_0,\dots,\underbrace{v_{i-1},v_{i+1}}_{\text{no }v_i\text{!}},\dots,v_p];
    $$
    \item $\partial_p$ is extended by linearity $\partial_p \sum_{i=1}^{l_p} f_i \sigma_i^p = \sum_{i=1}^{l_p} f_i \partial_p\sigma_i^p$;
    \item the boundary of the zero chain is zero.
\end{itemize}
\end{dfn}

Operator $\partial$ connects simplices of different dimensions. 
Now we need something to probe whether a simplex is ``internal'' to the polyhedron $|K|$ or faces ``ambient space''.
That will help us to ``define holes'' and the following substructures are exactly what is needed.

\begin{dfn}
\label{def_cycle}
\textbf{$p$-Cycles $Z_p(\mathcal{K})$} is a set of \textbf{$p$-cycles} $z_p$: $Z_p(\mathcal{K}) = \{z_p \in C_p(\mathcal{K}) | \partial_p z_p = 0\}$, i.e. the \textbf{kernel of $\partial_p$}.
\end{dfn}

\begin{dfn}
\label{def_boundary}
\textbf{$p$-Boundaries $B_p(\mathcal{K})$} is a set of \textbf{$p$-cycles} $b_p$: 
$$
B_p(\mathcal{K}) = \{b_p \in C_p(\mathcal{K}) | \exists c_{p+1} \in C_{p+1}(\mathcal{K}): \partial_{p+1} c_{p+1} = b_p\},
$$
i.e. the \textbf{image of $C_{p+1}(\mathcal{K})$ under $\partial_{p+1}$}.
\end{dfn}

\begin{thm}
$\partial_{p-1} \circ \partial_p = 0$.
Thus $B_p(\mathcal{K})\triangleleft Z_p(\mathcal{K})$.
\textnormal{$\blacktriangleleft$ see~\cite{maunder,nash}; see Definitions~\ref{def_cycle}, \ref{def_boundary}, note all subgroups of abelian groups are normal.$\blacksquare$}
\end{thm}

\begin{dfn}
\label{def_homology}
The \textbf{$p$-dimensional homology group} of $\mathcal{K}$ is the quotient group $H_p(\mathcal{K}) = Z_p(\mathcal{K}) / B_p(\mathcal{K})$.
\end{dfn}

Definition~\ref{def_cycle} rigorously defines cycles for us, while Definition~\ref{def_boundary} tells us which of them are ``filled in,'' i.e. contain no holes. The last Definition~\ref{def_homology} says ``consider closed cycles but disregard anything that is filled in,'' i.e. we are only interested in ``something with holes.'' Now the problem is that elements of $H_p(\mathcal{K})$ are not only those with one hole, but they are rather ``generated by holes.''
So we need to ``extract basis'' somehow and the following Theorems~\ref{thm_h_abelian} and~\ref{thm_decomp_gr} come in handy.

\begin{thm} 
\label{thm_h_abelian}
Homology group $H_p(\mathcal{K})$ of complex $\mathcal{K}$ is a finitely generated abelian group.
\textnormal{$\blacktriangleleft$ see~\cite{maunder,nash} $\blacksquare$}
\end{thm}

\begin{thm} 
\label{thm_decomp_gr}
Let $A$ be a finitely generated (not free!) abelian group with $n$ generators, then there exists a unique (except for the order of its members) list of primes $p_1$,...,$p_m$ (not necessarily distinct) and positive integers $s_1$,...,$s_m$, such that
$$
A \cong G \oplus \underbrace{\mathbb{Z}_{p_1^{s_1}} \oplus \cdots \oplus \mathbb{Z}_{p_m^{s_m}}}_{T},
$$
where $T$ is called the \textbf{torsion subgroup}, $\mathbb{Z}_{{p_i^{s_i}}}$ are cyclic groups of order $p_i^{s_i}$, and $G$ is free abelian group. 
The rank of $G$ is $n - m$. 
\textnormal{$\blacktriangleleft$ see~\cite{hungerford} Theorem 2.6$\blacksquare$}
\end{thm}

The procedure is somewhat similar to the decomposition of a number into prime factors.
In practice, it is performed by representing operators $\partial_p$ as matrices and employing the Smith normal form~\cite{smith}, but here we only outline the theoretical basis.

\begin{dfn}
\label{def_betti}
The rank of $G$ from Theorem~\ref{thm_decomp_gr} for $A = H_p(\mathcal{K})$ is called the \textbf{$p$-th Betti number $\beta_p$} of the geometric simplicial complex $K$. 
\end{dfn}

Please note: despite the fact we have defined Betti numbers for abstract simplicial complex $\mathcal{K}$, they are inherently connected to its geometric realization $K$.
Thus Betti numbers can be treated as topological characteristics of the polyhedron $|K|$.
Moreover, topology makes no distinction between homeomorphic spaces, thus the same characteristic can be prescribed to any space $\mathbb{X}$ that is homeomorphic to $|K|$.
This property is summarized by the following.

\begin{dfn}
\label{def_hom_sc}
A \textbf{triangulation} of topological space $\mathbb{X}$ is a geometric simplicial complex $K$ together with a homeomorphism $f: |K|\to \mathbb{X}$.
If there exists such $K$ the space $\mathbb{X}$ is called \textbf{triangulable}.
The homotopy groups of a triangulable space $\mathbb{X}$ are defined $H_p(\mathbb{X}) = H_p(\mathcal{K})$.
\end{dfn}

\begin{thm}
Homotopy groups $H_p(\mathbb{X})$ and $H_p(\mathbb{Y})$ of homeomorphic topological spaces are isomorphic for each $p$.
\textnormal{$\blacktriangleleft$ see~\cite{vick} Theorem 1.7$\blacksquare$}
\end{thm}

The latter means that homotopy groups of the triangulable space (Definition~\ref{def_hom_sc}) are well-defined and that the notion of Betti numbers can be extended to topological spaces that are homeomorphic to some polyhedron $|K|$\footnote{For visuals please check \url{https://fbeilstein.github.io/topological_data_analysis/homology_explorer/homology_explorer.html}}.

\subsection{From Data to Simplicial Complexes}
At the moment we have done nothing to our dataset and it's about time to take the data points into account. Here we present a few methods of creating simplicial complexes out of date apoints thus connecting their positions to the topology of a certain manifold. There are different methods of constructing an abstract simplicial complex from the data points and the exact choice may depend on the problem and computational resources at your disposition. Note that Theorem~\ref{thm_simplex_real} warranties that whatever we come up with will have a geometric representation, but its dimension may be higher than the original dataset we started with.
Here we start with the two most popular choices ($\alpha$ complex that we use in the article can be thought of as a variation of \u{C}ech complex)

\begin{dfn}
\label{def_ripps}
Let $(M;\rho)$ be a metric space.
Given a finite set of points $x_i$ in $M$ (the dataset) and a real number $\alpha > 0$, the (abstract) \textbf{Vietoris-Rips complex} is constructed as follows:
\begin{itemize}
\item its abstract vertices $v_i$ are in one-to-one correspondence with $x_i$ from $M$;
\item it contains a simplex $\sigma^n = (v_0;\cdots;v_n)$ if and only if for each pair of vertices $v_i$ and $v_j$ the distance between corresponding points in $M$ is $\rho(x_i;x_j) \leq \alpha$.
\end{itemize}
\end{dfn}

\begin{dfn}
\label{def_cech}
Let $(M;\rho)$ be a metric space.
Given a finite set of points $x_i$ in $M$ (the dataset) and a real number $\alpha > 0$, the (abstract) \textbf{\u{C}ech complex} is constructed as follows:
\begin{itemize}
\item its abstract vertices $v_i$ are in one-to-one correspondence with $x_i$ from $M$;
\item it contains a simplex $\sigma^n = (v_0;\cdots;v_n)$ if and only if there is non-empty intersection $\bigcap_{i=0}^n B(x_i;\alpha) \neq \emptyset$.
\end{itemize}
\end{dfn}

Computer scientists often prefer the Vietoris-Rips complex (Definition~\ref{def_ripps}) as you need less computational resources to calculate it in higher-dimensional spaces.
On the other hand, the following Theorem~\ref{thm_nerve} is a cornerstone of topological data analysis that connects \u{C}ech complex and topology of the union of balls from Definition~\ref{def_cech}, thus it's often preferred by physicists.

\begin{dfn}
Given an open cover $\mathcal{U} = (U_i)_{i \in I}$ of topological space $\mathbb{X}$, the nerve of $\mathcal{U}$ is the abstract simplicial complex $C(\mathcal{U})$ whose vertices are the $U_i$'s and such that
$$
\sigma = (U_{i_0};\cdots;U_{i_k}) \in C(\mathcal{U}) \iff \bigcap_{j=0}^k U_{i_j} \neq \emptyset.
$$
\end{dfn}

\begin{thm}
\label{thm_nerve}
(Nerve Theorem). Let $\mathcal{U} = (U_i)_{i \in I}$ be a cover of a paracompact space $\mathbb{X}$ by open sets such that the intersection of any subcollection of the $U_i$'s is either empty or contractible. Then, $\mathbb{X}$ and the nerve $C(\mathcal{U})$ are homotopy equivalent.
\textnormal{$\blacktriangleleft$ see~\cite{hatcher} Corollary 4G.3 or~\cite{alexandroff}$\blacksquare$}
\end{thm}

The latter Theorem~\ref{thm_nerve} is very famous but its statement needs few more complex concepts from topology such as ``paracompactness,'' ``contractibility,'' ``homotopy equivalence.''
To avoid the problem let's reformulate it in a more convenient form.

\begin{thm}
\label{thm_cech_to_balls}
Assume that we are given a finite set of points $x_i$ in $\mathbb{R}^n$ and a real number $\alpha > 0$.
Consider homotopy groups $H_p(\mathbb{X})$ of the topological space $\mathbb{X}$ obtained as a union of closed balls $B(x_i;\alpha)$.
$H_p(\mathbb{X})$ are isomorphic to the homotopy groups of the polyhedron of the realization of \u{C}ech complex of these points with parameter $\alpha$.
\textnormal{$\blacktriangleleft$ This is a partial case of \cite{borsuk} Corollary 3, p.234 $\blacksquare$}
\end{thm}

Please note that the polyhedron of \u{C}ech complex may be non-homeomorphic to $\mathbb{X}$ even in simple cases and belong to a higher-dimensional space\footnote{For visuals please check \url{https://fbeilstein.github.io/topological_data_analysis/persistent_homology_explorer/persistent_homology_explorer.html}}. 
Also note, that everything in the Definitions~\ref{def_cech} and~\ref{def_ripps} depends on $\alpha$: the Betti curves we draw in the article depend on this $\alpha$. A manifold is not something given to us but rather hypothesized by us, thus in practice, it may be reasonable to use complexes other than \u{C}ech complex, especially when they are easier to compute. In this work, we used $\alpha$-complex that is homotopy equivalent to the \u{C}ech complex, but it could have been some different complex as well. 

\begin{dfn}
\label{def_betti_curve}
The \textbf{$p$-th Betti curve} is a plot of $\beta_p$ from Definition~\ref{def_betti} vs parameter $\alpha$ that we used to construct a simplicial complex (see Definitions~\ref{def_cech} and~\ref{def_ripps}).
\end{dfn}

The last Definition~\ref{def_betti_curve}, basically, finishes the consideration.


\begin{thebibliography}{3}
{\small
\bibitem{Hamilton86} A. J. S. Hamilton, J. R. Gott, D. Weinberg. The Astrophysical Journal, \textbf{309}(1), 1-12 (1986); \href{https://doi.org/10.1086/164571}{https://doi.org/10.1086/164571}
\bibitem{Sahni98} V. Sahni, B.S. Sathyaprakash, S.F. Shandarin, The Astrophysical Journal, \textbf{495}(1), L5–L8 (1998); \href{https://doi.org/10.1086/311214}{https://doi.org/10.1086/311214}
\bibitem{Sheth03} J.V. Sheth, V. Sahni, S.F. Shandarin, B.S. Sathyaprakash, Monthly Notices of the Royal Astronomical Society, \textbf{343}(1), 22–46 (2003); \href{https://doi.org/10.1046/j.1365-8711.2003.06642.x}{https://doi.org/10.1046/j.1365-8711.2003.06642.x}
\bibitem{Sousbie1} T. Sousbie, Monthly Notices of the Royal Astronomical Society, \textbf{414}(1), 350–383 (2011); \href{https://doi.org/10.1111/j.1365-2966.2011.18394.x}{https://doi.org/10.1111/j.1365-2966.2011.18394.x}
\bibitem{Sousbie2} T. Sousbie, C. Pichon, H. Kawahara, Monthly Notices of the Royal Astronomical Society, \textbf{414}(1), 384–403 (2011); \href{https://doi.org/10.1111/j.1365-2966.2011.18395.x}{https://doi.org/10.1111/j.1365-2966.2011.18395.x}
\bibitem{Feldbrugge19} J. Feldbrugge, M. van Engelen, R. van de Weygaert, P. Pranav, G. Vegter, J. Cosmol. Astropart. Phys. \textbf{201} (09)}, 052–052 (2019); \href{https://doi.org/10.1088/1475-7516/2019/09/052}{https://doi.org/10.1088/1475-7516/2019/09/052}
\bibitem{Park13} C. Park et al., Journal of The Korean Astronomical Society \textbf{46}(3), 125–131 (2013); \href{https://doi.org/10.5303/JKAS.2013.46.3.125}{https://doi.org/10.5303/JKAS.2013.46.3.125}
\bibitem{Giri21}S.K. Giri, G. Mellema, Monthly Notices of the Royal Astronomical Society \textbf{505}(2), 1863–1877 (2021); \href{https://doi.org/10.1093/mnras/stab1320}{https://doi.org/10.1093/mnras/stab1320}
\bibitem{Elbers19} W. Elbers, R. van de Weygaert, Monthly Notices of the Royal Astronomical Society, \textbf{482}(2), 1523–1538 (2019); \href{https://doi.org/10.1093/mnras/stz908}{https://doi.org/10.1093/mnras/stz908}
\bibitem{Elbers23} W. Elbers, and R. van de Weygaert, “Persistent topology of the reionization bubble network – II. Evolution and classification,” Monthly Notices of the Royal Astronomical Society \textbf{520} (2), 2709–-2726 (2023); \href{https://doi.org/10.1093/mnras/stad120}{https://doi.org/10.1093/mnras/stad120}
\bibitem{Cole18} A. Cole, and G. Shiu, Journal of Cosmology and Astroparticle Physics \textbf{2018(03)}, 025–025 (2018); \href{https://doi.org/10.1088/1475-7516/2018/03/025}{https://doi.org/10.1088/1475-7516/2018/03/025}
\bibitem{Xu19} X. Xu, J. Cisewski-Kehe, S.B. Green, D. Nagai, Astronomy and Computing \textbf{27}, 34–52 (2019); \href{https://doi.org/10.1016/j.ascom.2019.02.003}{https://doi.org/10.1016/j.ascom.2019.02.003}
\bibitem{Heydenreich21} S. Heydenreich, B. Brück, and J. Harnois-Déraps, Astronomy and Astrophysics, \textbf{648}, A74 (2021); \href{https://doi.org/10.1051/0004-6361/202039048}{	https://doi.org/10.1051/0004-6361/202039048}
\bibitem{Codis18} S.Codis, D. Pogosyan, C. Pichon, Monthly Notices of the Royal Astronomical Society, \textbf{479}(1), 973–993 (2018); \href{https://doi.org/10.1093/mnras/sty1643}{https://doi.org/10.1093/mnras/sty1643}
\bibitem{Shivashankar16} N. Shivashankar, P. Pranav, V. Natarajan, R. van de Weygaert, E.G.P. Bos, S. Rieder, IEEE Transactions on Visualization and Computer Graphics, \textbf{22}(6), 1745–1759 (2016); \href{https://doi.org/10.1109/TVCG.2015.2452919}{https://doi.org/10.1109/TVCG.2015.2452919}
\bibitem{Pranav16} P. Pranav, H. Edelsbrunner, R. van de Weygaert, G. Vegter, M. Kerber, B.J.T. Jones, and M. Wintraecken, Monthly Notices of the Royal Astronomical Society, \textbf{465}(4), 4281–4310 (2016); \href{https://doi.org/10.1093/mnras/stw2862}{https://doi.org/10.1093/mnras/stw2862}
\bibitem{Weygaert11} R. van de Weygaert et al., arxiv:1110.5528 (2011); \href{https://doi.org/10.48550/arXiv.1110.5528}{https://doi.org/10.48550/arXiv.1110.5528}
\bibitem{Weygaert13} R. van de Weygaert et al. Transactions on Computational Science XIV. Lecture Notes in Computer Science, vol 6970. Springer, Berlin, Heidelberg (2011); \href{https://doi.org/10.1007/978-3-642-25249-5_3}{https://doi.org/10.1007/978-3-642-25249-53}
\bibitem{Wilding21} G. Wilding, K. Nevenzeel, R. van de Weygaert, G. Vegter, P. Pranav, B.J.T. Jones, K. Efstathiou, J. Feldbrugge, Monthly Notices of the Royal Astronomical Society \textbf{507}(2), 2968–2990 (2021); \href{https://doi.org/10.1093/mnras/stab2326}{https://doi.org/10.1093/mnras/stab2326}
\bibitem{Bermejo22} R. Bermejo, G. Wilding, R. van de Weygaert, B.J.T. Jones, G. Vegter, K. Efstathiou, Monthly Notices of the Royal Astronomical Society, \textbf{529}(4), 4325–4353 (2022); \href{https://doi.org/10.1093/mnras/stae543}{https://doi.org/10.1093/mnras/stae543}
\bibitem{Ouellette23}A. Ouellette, G. Holder, and E. Kerman, Monthly Notices of the Royal Astronomical Society, \textbf{523}(4), 5738–5747 (2023); \href{https://doi.org/10.1093/mnras/stad1765}{https://doi.org/10.1093/mnras/stad1765}
\bibitem{Tsizh23}M. Tsizh, V. Tymchyshyn, F. Vazza, Monthly Notices of the Royal Astronomical Society, \textbf{522}(2), 2697–2706 (2023); \href{https://doi.org/10.1093/mnras/stad1121}{https://doi.org/10.1093/mnras/stad1121}
\bibitem{alexandroff1928} P. Alexandroff. Mathematische Annalen, \textbf{98} (1), 617-635 (1928). 
\bibitem{Edelsbrunner02} H. Edelsbrunner, D. Letscher, A. Zomorodian, Discrete and Computational Geometry, \textbf{28} (4), 511–533 (2002);  \href{https://doi.org/10.1007/s00454-002-2885-2}{https://doi.org/10.1007/s00454-002-2885-2}
\bibitem{zomorodian09} A. Zomorodian, \textit{Topology for Computing} (Cambridge Monographs on Applied and Computational Mathematics, Series Number 16, Cambridge Univ Press, 2009).
\bibitem{zomorodian12} A. Zomorodian, Advances in Applied and Computational Topology (Proceedings of Symposia in Applied Mathematics, \textbf{70}), 1–39 (2012); \href{https://doi.org/10.1090/psapm/070}{https://doi.org/10.1090/psapm/070}
\bibitem{carlsson04} A. Zomorodian, G. Carlsson, Discrete and Computational Geometry, \textbf{33}, 249–274, (2005); \href{https://doi.org/10.1007/s00454-004-1146-y}{https://doi.org/10.1007/s00454-004-1146-y}
\bibitem{carlsson09} G. Carlsson, The Bulletin of American Mathematical Society, \textbf{46}(2), 255-308, (2009); \href{https://doi.org/10.1090/S0273-0979-09-01249-X}{https://doi.org/10.1090/S0273-0979-09-01249-X}
\bibitem{Wasserman18} L. Wasserman, Annual Review of Statistics and Its Application, \textbf{5}(1), 501–532 (2018); \href{https://doi.org/10.1146/annurev-statistics-031017-100045}{https://doi.org/10.1146/annurev-statistics-031017-100045} 
\bibitem{bolshoi} A. Klypin, S. Trujillo-Gomez, J. Primack, The Astrophysical Journal, \textbf{740} (2), 102 (2011); \href{https://doi.org/10.1088/0004-637X/740/2/102}{https://doi.org/10.1088/0004-637X/740/2/102} 
\bibitem{multidark} F. Prada et al.  Monthly Notices of the Royal Astronomical Society \textbf{423}(4), 3018-3030 (2012); \href{https://doi.org/10.1111/j.1365-2966.2012.21007.x}{https://doi.org/10.1111/j.1365-2966.2012.21007.x} 
\bibitem{Giocoli:2018gqh} C.~Giocoli, M.~Baldi and L.~Moscardini, Monthly Notices of the Royal Astronomical Society, \textbf{481} (2), 2813-2828 (2018); \href{https://doi.org/10.1093/mnras/sty2465}{https://doi.org/10.1093/mnras/sty2465}
\bibitem{Lukic07} Z. Lukic, K. Heitmann, S. Habib, S. Bashinsky, P. M. Ricker, The Astrophysical Journal, \textbf{671} (2), 1160 (2007); \href{https://doi.org/10.1086/523083}{https://doi.org/10.1086/523083}
\bibitem{wmap5} E. Komatsu1 et al., The Astrophysical Journal Supplement, \textbf{180} (2), p.330-376 (2009); \href{https://doi.org/10.1088/0067-0049/180/2/330}{https://doi.org/10.1088/0067-0049/180/2/330}
\bibitem{planck15} Planck Collaboration, Astronomy and Astrophysics, \textbf{594}, October 2016, A13 (2016); \href{https://doi.org/10.1051/0004-6361/201525830}{https://doi.org/10.1051/0004-6361/201525830}
\bibitem{Santos19} F.A.N. Santos, E.P. Raposo, M.D. Coutinho-Filho, M. Copelli, C.J. Stam, and L. Douw, Phys. Rev. E \textbf{100}(3), 032414 (2012); \href{https://doi.org/PhysRevE.100.032414}{https://doi.org/PhysRevE.100.032414}
\bibitem{Euclid:2023uha} Euclid Collaboration, Astronomy and Astrophysics, \textbf{675}, A120 (2023); \href{https://doi.org/10.1051/0004-6361/202346017}{https://doi.org/10.1051/0004-6361/202346017}
\bibitem{Gluzberg2023} V.E Gluzberg, Y.A. Katz, Communications in Nonlinear Science and Numerical Simulation, \textbf{121}, 107216 (2023); \href{https://doi.org/10.1016/j.cnsns.2023.107216}{https://doi.org/10.1016/j.cnsns.2023.107216}
\bibitem{maunder} C. R. F. Maunder. \textit{Algebraic topology} (Courier Corporation, 1996).
\bibitem{nash} C. Nash, S. Sen. \textit{Topology and geometry for physicists} (Elsevier, 1988). 
\bibitem{chasal} Chazal, Frédéric, and Bertrand Michel. Frontiers in artificial intelligence \textbf{4}, 108 (2021); \href{https://doi.org/10.3389/frai.2021.667963}{https://doi.org/10.3389/frai.2021.667963}
\bibitem{hatcher} A. Hatcher. \textit{Algebraic topology}, (Cambridge University Press, 2002).
\bibitem{hungerford} T. W. Hungerford. \textit{Algebra. Vol. 73.} (Springer Science \& Business Media, 2012).
\bibitem{alexandroff} P. Alexandroff,  Mathematische Annalen \textbf{98} (1), 617-635 (1928).
\bibitem{smith} H. Smith, Philosophical transactions of the royal society of London, \textbf{151}, 293-326 (1861).
\bibitem{borsuk} K. Borsuk. Fundamenta Mathematicae, \textbf{35.1}, 217-234 (1948).
\bibitem{vick} J. W. Vick,  \textit{Homology theory: Homology Theory} (Academic Press, 1973).
\bibitem{Landy93} S.D. Landy, A.S. Szalay, The Astrophysical Journal, \textbf{412}, 64 (1993); \href{https://doi.org/10.1086/172900}{https://doi.org/10.1086/172900}

\end{thebibliography}
\end{document}